\documentclass[aps,prl,reprint,twocolumn,superscriptaddress]{revtex4-2}
\usepackage{graphicx} 
\usepackage{amsmath,amsthm, amssymb}
\usepackage{physics}
\usepackage{xcolor}
\usepackage{time}
\usepackage[pdftex,colorlinks=true]{hyperref}
\usepackage[colorinlistoftodos]{todonotes}
\usepackage{sidecap}
\usepackage{amsfonts}
\usepackage{mathtools}
\usepackage{bm}
\usepackage{cleveref}
\usepackage{color,soul}
\usepackage{todonotes}
\usepackage{ragged2e}
\newcounter{todocounter}

\def \be{\begin{equation}}
\def \ee{\end{equation}}
\def \bea{\begin{eqnarray}}
\def \eea{\end{eqnarray}}

\newcommand{\rl}{\{\bm{r}; \bm{l}\}}

\begin{document}

\title{Solving fractional electron states in twisted MoTe$_2$ with deep neural network}
\author{Di Luo}
\affiliation{Department of Electrical and Computer Engineering, University of California, Los Angeles, California 90095, USA}
\affiliation{Department of Physics, Massachusetts Institute of Technology, Cambridge, Massachusetts 02139, USA}
\affiliation{The NSF AI Institute for Artificial Intelligence and Fundamental Interactions}
\affiliation{Department of Physics, Harvard University, Cambridge, MA 02138, USA}
\author{Timothy Zaklama}
\affiliation{Department of Physics, Massachusetts Institute of Technology, Cambridge, Massachusetts 02139, USA}
\author{Liang Fu}
\affiliation{Department of Physics, Massachusetts Institute of Technology, Cambridge, Massachusetts 02139, USA}

\date{\today}

\begin{abstract}
The emergence of moiré materials, such as twisted transition-metal dichalcogenides (TMDs), has created a fertile ground for discovering novel quantum phases of matter. However, solving many-electron problems in moir\'e systems presents significant challenges due to strong electron correlation and strong moir\'e band mixing. Recent advancements in neural quantum states hold the promise for accurate and unbiased variational solutions. Here, we introduce a powerful neural wavefunction to solve ground states of twisted MoTe$_2$ across various fractional fillings, reaching unprecedented accuracy and system size. From the full structure factor and quantum weight, we conclude that our neural wavefunction accurately captures both the electron crystal at $\nu=1/3$ and various fractional quantum liquids in a unified manner.   
\end{abstract}

\maketitle

{\it Introduction---.} Many-electron physics in moir\'e materials has become an exciting frontier of condensed matter research due to the discovery of novel quantum phases of matter. Thanks to the unprecedented level of tunability,   transition metal dichalconegide (TMD) moir\'e superlattices host a wide array of correlated and topological electronic states, including generalized Wigner crystals \cite{regan2020mott}, superconductivity \cite{xia2024unconventional,guo2024superconductivity}, quantum anomalous Hall states \cite{li2021quantum}, and fractional quantum anomalous Hall states \cite{cai2023signatures, park2023observation, zeng2023thermodynamic, xu2023observation}. Remarkably, all these diverse phenomena emerge from two-dimensional electron gas modulated by the moir\'e superlattice, therefore allowing for a unified theoretical description.

Despite the conceptual simplicity, solving many-electron problems in moiré semiconductors presents significant challenges due to the lack of accurate and unbiased methods for simulating interacting electron Hamiltonian in continuous space. The Hartree-Fock method misses electron correlations.  The exact diagonalization method is not only limited to small system size, but also relies on truncation to one or few lowest moir\'e bands \cite{morales2021metal, li2021spontaneous, crepel2023anomalous, reddy2023fractional, wang2023fractional}. However, in many moir\'e systems, the characteristic Coulomb interaction is (considerably) larger than the moir\'e band gap, leading to strong band mixing \cite{zhang2020moire, reddy2023artificial}. Indeed, band-projected exact diagonalization finds that increasing the number of included moir\'e bands can significantly lower the total energy~\cite{abouelkomsan2024band, yu2024fractional}. Clearly, developing accurate and efficient numerical methods is crucial for advancing precision many-body theory in moiré materials and quantitatively understanding or predicting experimental observables.

Recent advancements in artificial intelligence, particularly in machine learning and neural network quantum states~\cite{Carleo602}, have opened new possibilities for solving many-body problems~\cite{chen2023autoregressive,robledo2022fermionic, chen2022simulating, doi:10.1126/science.aag2302, Hibat_Allah_2020, PhysRevLett.124.020503, Irikura_2020, PhysRevResearch.3.023095, Han_2020,ferminet,Choo_2019,rnn_wavefunction,paulinet,Glasser_2018,Stokes_2020,Nomura_2017,martyn2022variational,Luo_2019,PhysRevLett.127.276402, https://doi.org/10.48550/arxiv.2101.07243,luo2022gauge}. Neural networks have demonstrated remarkable power in representing strongly correlated and highly entangled quantum states \cite{ferminet, paulinet, glasser2018neural, pescia2022neural}. While early applications of neural networks focused on quantum chemistry and lattice models, this approach has recently been developed to solve continuum models~\cite{cassella2023discovering,pescia2023message,kim2023neural,lou2023neural,entwistle2023electronic,wilson2022wave,li2022ab,scherbela2022solving,adams2021variational,smith2024ground,luo2023pairing}, including fractional quantum Hall systems \cite{teng2024solving, qian2024taming} and semiconductor heterobilayers~\cite{li2024emergent,Luo2024NNMoire, geier2025attention}.

In this work, we introduce a powerful neural network (NN) wavefunction ansatz to variationally solve the continuum Hamiltonian of interacting electrons in twisted TMD homobilayers at fractional moir\'e fillings. 
Working with the full Hilbert space without band trunctation, our NN method substantially outperforms band-projected exact diagonalization that includes as many as 5 bands. The expressive power of our NN architecture enables us to obtain distinct ground states across many filling factors---including generalized Wigner crystal, Fermi liquid and a sequence of fractional quantum anomalous Hall (FQAH) liquids---in a unified and unbiased manner.

\begin{figure}[t]
    \centering
    \includegraphics[width=0.7\columnwidth]{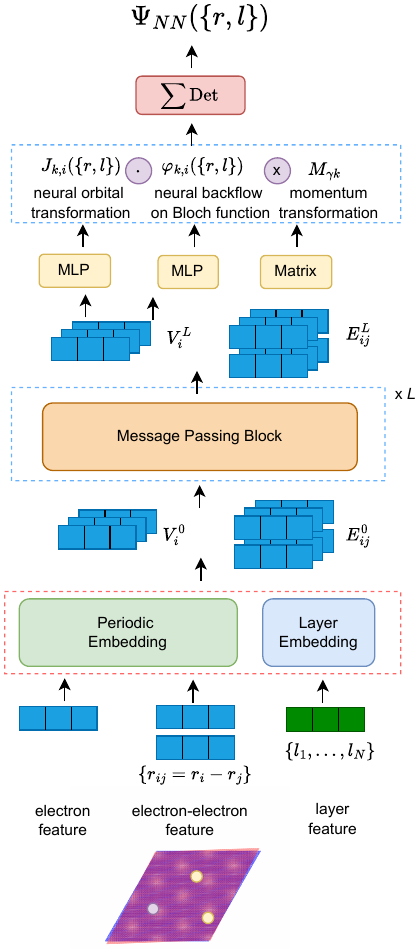}
    \caption{Many-body neural Bloch wavefunction schematic.}
    \label{fig:nn}
\end{figure}

{\it Interacting continuum model---.} We study the following interacting continuum model for holes in $t$MoTe$_2$:   
\begin{align}\label{eq:Ham}
    \begin{split}
        \hat{H}&= \sum_{i}\left[ \frac{\bm{p}_i^2}{2m}+\bm{J}(\bm{r}_i)\cdot\bm{\sigma}_i + V(\bm{r}_i)\right] \\
        &+ \frac{1}{2}\sum_{i}\sum_{j\neq i}U(\bm{r}_i-\bm{r}_j) 
    \end{split}
\end{align}
where the Pauli matrices $\bm \sigma$ are associated with the layer pseudospin. The first term is the one-body Hamiltonian \cite{wu2019topological}, including kinetic energy, interlayer tunneling $J_{\pm}=J_x \pm i J_y$ and intra-layer moir\'e potential $V_l = V\pm J_z$ ($l=1,2$ denotes the layer). Explicitly, $J_-(\bm{r}) = w(e^{i\bm{Q}_1\cdot \bm{r}}+e^{i\bm{Q}_2\cdot \bm{r}}+e^{i\bm{Q}_3\cdot \bm{r}})$ and $V_{1,2}(\bm{r}) = -2V_0\sum_{i=1}^{3}\cos(\bm{g}_i\cdot\bm{r} \pm \phi)$ are spatially varying with the moir\'e periodicity, where $\bm{Q}_{1,2,3}$ denote the lowest-order reciprocal vectors related by three-fold rotation; $V_0$ and $\phi$ characterize the moir\'e potential depth and landscape.  

Our numerical calculation is performed on finite size clusters with periodic boundary conditions. 
For the two-body interaction, we use a Coulomb interaction adapted to periodic boundary conditions, $U(\bm{r}) = \sum_{n,m\in \mathbb{Z}}e^2/(\epsilon|\bm{r}+n\bm{L}_1+m\bm{L}_2|)$ \cite{wu2013bloch, reddy2023fractional, geier2025attention}. 
We use the following continuum model parameters for twisted tMoTe$_2$~\cite{reddy2023fractional}: $(V,w,\phi,m^*) = \left(11.2 \text{meV}, 13.3 \text{meV}, -91^{\circ},0.62 m_e\right)$.

{\it Neural Bloch wavefunction---.} Our neural wavefunction  for $N$ electrons on two layers takes the form of a generalized Slater determinant: 
\begin{equation}
    \bm{\Psi}_{NN}(\rl)  
    =\sum_w \det[\Phi^{w}_{\gamma, i}  (\rl)] 
    \label{MBNBW}
\end{equation}
where $\Phi_{\gamma, i} (\rl)$ denotes a generalized orbital $\gamma$ that depends on the location $\bm{r}$ and layer index $\bm{l}$ of all  particles $\rl=(r_1, \dots, r_N; l_1, \dots, l_N)$ \cite{Luo_2019, ferminet,paulinet}.  In the special case $\Phi_{\gamma, i} (\rl) = \Phi_{\gamma} (r_i, l_i)$ is a one-particle wavefunction, our wavefunction reduces to a sum of standard Slater determinants indexed by $w$. 
For simplicity, we omit the symbol $w$ in the following discussion since the properties in our construction applies to each generalized orbital. 

The expressive power of our neural wavefunction comes from the fact that $\Phi_{\gamma, i} (\rl)$ is a many-body generalization of Bloch states constructed by the neural network architecture shown in Fig~\ref{fig:nn}: 
\begin{equation}
    \Phi_{\gamma, i} (\rl)  = \sum_{k} M_{\gamma k} J_{k, i}(\rl) \varphi_{k, i} (\rl). 
    \label{MBNBW}
\end{equation}
where $M_{\gamma k}$ is a momentum transformation matrix, $J_k$ is neural orbital transformation, and $\varphi_k $ is neural network backflow from Bloch function, as we elaborate below. We note that $M_{\gamma k}$, $J_k(\rl)$ and $\varphi_{k,i}(\rl)$ have different variational parameters in each generalized orbital when multiple determinants are used in Eq.~\ref{MBNBW}.

The neural network backflow transformation is designed to transform the single particle orbital into a many-body orbital. 
To start, we consider a set of one-body Bloch basis functions at momentum $k$,

\begin{equation}
\varphi_{k}(r_i,l_i) = \sum_{g} u^{l_i}_{k g}e^{i(k+g)r_i}
\end{equation}
where $g$ is the reciprocal lattice momentum respectively, and $u_{kg}^{l_i}$ is the Bloch function coefficient. 
To promote the one-body Bloch basis to the many-body neural Bloch wavefunction, we first introduce a backflow transformation in position space
\begin{equation}
r_i \rightarrow \bm{r_i} = f_i(\rl). 
\end{equation}
Specifically, the backflow transformation $f_i$ is implemented by a powerful graph neural network, message passing neural network (MPNN), which has been recently used in neural wavefunctions 
~\cite{pescia2023message,kim2023neural,smith2024ground,luo2023pairing, Luo2024NNMoire}. 
MPNN starts with a graph representation of the particle configurations as a graph $G_0 = {V^0,E^0}$, with node feature $V^{0}_i$ and edge feature $E^{0}_{ij}$. 
To satisfy the periodic condition and lattice translation symmetry, we use $(\text{sin}(g r_i),\text{cos}(g r_i))$ with reciprocal momentum $g$ as initial feature for $V^{0}_{i}$, and $(\text{sin}(k (r_i - r_j)),\text{cos}(k (r_i - r_j)))$ with mesh momentum $k$ as initial feature for $E^{0}_{ij}$ (see Supplementary Materials for details). With the initial graph features, MPNN iteratively updates the graph a certain number of times, $t$, until a new set of graph features $V^{t}_i$ and $E^{t}_{ij}$ are reached. The physics motivation for message passing neural network is to capture the correlation among particles through iteration while preserving the permutation equivariance of the output. Notice that $V^{t}_i$ is a function of the many-body configuration $\bm{r}$. The neural network backflow is then given by $\bm{r}_i = r_i + WV^{t}_i$, where $W$ is a complex-valued matrix. This procedure promotes the one-body orbital 
into a many-body correlated orbital $\varphi_{k,i}(\rl)$. The details of the
message passing network structure is
provided in the Supplementary Materials.

\begin{figure}[t]
    \includegraphics[width=.45\textwidth]{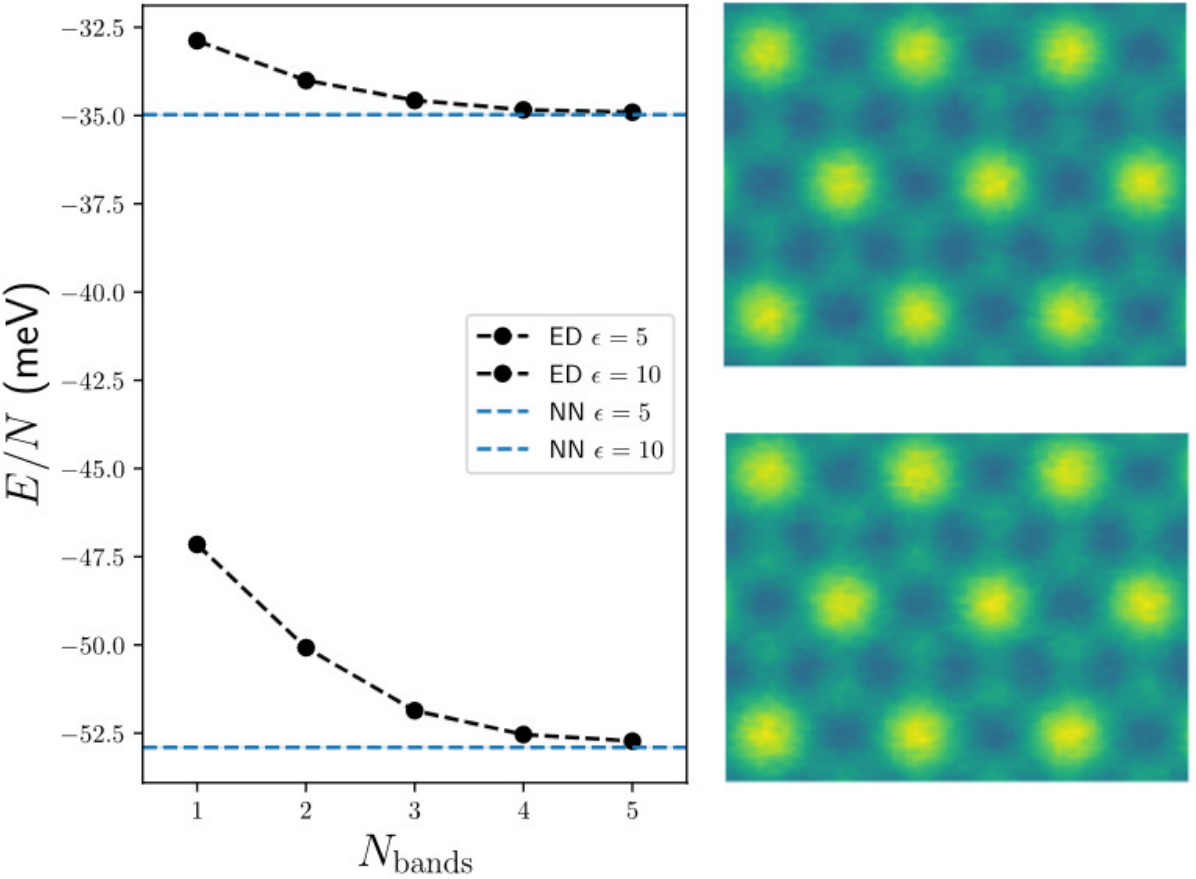} 
    \caption{(Left) Ground state energies from multiband ED in comparison with NN at $\theta=3.0$ for $\nu=2/3$ in a 3x3 cluster. Statistical error bar on NN is included but negligible. (Right) Ground state charge density from NN at $\epsilon=5$ (top) and $10$ (bottom).}
    \label{Fig3: eps}
\end{figure}

Next, we introduce a neural orbital transformation to enhance the expressivity of the neural network backflow orbital. 
This is achieved by multiplying $\varphi_{k,i}(\rl)$ with $J_{k,i}(\rl)=\text{MLP}(\{k\},V_i^t)$, where MLP is a two-layer complex-valued fully connected neural network. We note that the complex function nature of our neural orbital transformation is crucial since it allows for a change in the density and the nodal structure of the wavefunction directly, which is more general than the conventional real-valued Jastrow function.

The generalized orbitals obtained after backflow and orbital transformations remain many-body momentum eigenstates. 
To further enhance the representation power of the many-body neural Bloch wavefunction, we introduce the momentum
transformation matrix $M_{\lambda k}$, where the $\lambda=1,...,N$ with $N$ the number of particles and $k$ is the momentum index. Note that the number of $k$ points can be greater than $N$.   
This procedure can mix states with different many-body momenta to increase the representation power of the neural network, which can be seen from the Cauchy-Binet formula:  

\begin{equation}
    \det(MQ) = \sum_{S \subseteq \{1, 2, \dots, k\}, |S| = N} \det(M_S) \det(Q_S)
\end{equation}
where $M$ is the momentum transformation matrix $M_{\gamma k}$, and $Q=J \odot \varphi$ is the elementwise product of the neural orbital transformation $J_{k,i}$, with the neural backflow $\varphi_{k,i}$. From the above formula, even with only one determinant, the momentum transformation induces a sum of determinants, which offers a more powerful and efficient representation of the neural wavefunction.

\begin{figure}[b]
    \includegraphics[width=.45\textwidth]{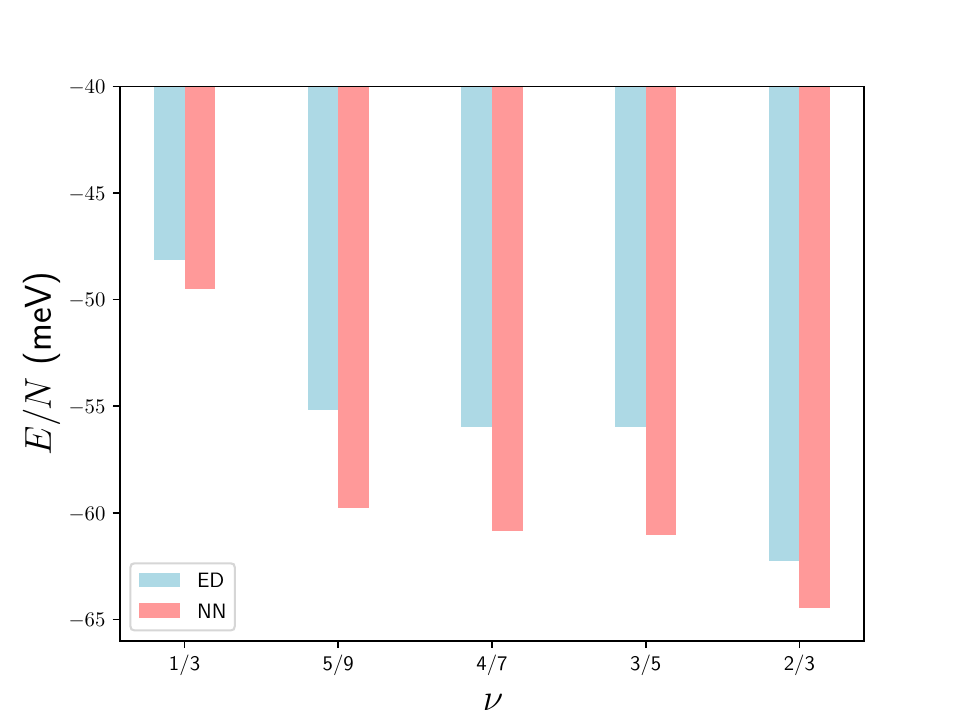} 
    \caption{Ground state energies from 1-band ED and NN at $\nu=\frac{1}{3}$$,\frac{5}{9}$$,\frac{4}{7}$$,\frac{3}{5}$$,\frac{2}{3}$ at $\theta=3.0$. Calculations are performed on clusters with 27, 27, 28, 25, 27 unit cells respectively.}
    \label{Fig2: filling}
\end{figure}

Our neural network construction guarantees that $\Phi_{\gamma, i} (\rl)$ is permutation equivariant, i.e. for any $i,j$ with $l_i=l_j$, 
\begin{equation}
 \Phi_{\gamma, i} (P_{ij}(\rl))  = \Phi_{\gamma, j}(\rl)  
\end{equation}
where $P_{ij}(\rl)=(\dots,r_j,\dots,r_i, \dots ; \bm{l}) 
$ interchanges the $i$- and $j$-th arguments in $\mathbf{r}$. 

The permutation equivariant property of the generalized orbital, together with the anti-symmetry property of the determinant, guarantees the anti-symmetry of the neural wavefunction $\Psi_{NN}$. 

We also note that an innovation of our architecture is its flexibility to handle layer degree of freedom directly, a feature absent in previous neural network wave functions. Both the Bloch basis function and neural network transformation incorporate the layer pseduospin, enabling the representation of general wavefunctions without layer particle number conservation. 

\begin{figure*}[t]
    \centering
    \includegraphics[width=0.9\textwidth]{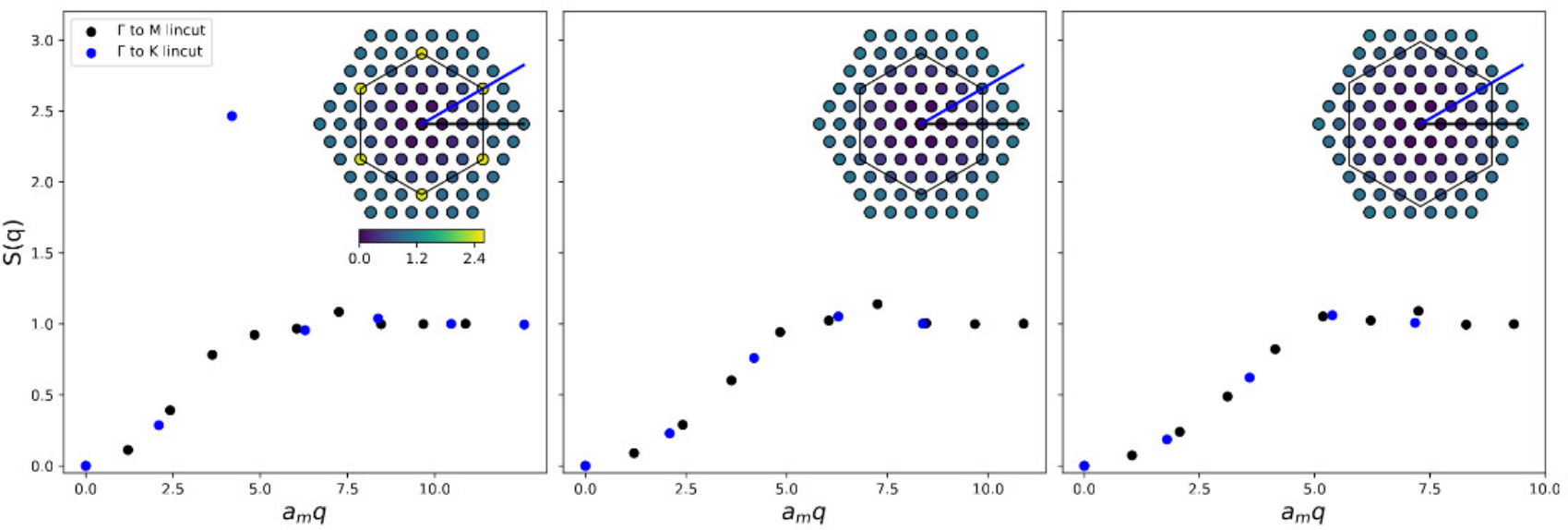} 
    \caption{Full structure factor at $\nu=\frac{1}{3}$,$\frac{2}{3}$,$\frac{4}{7}$ at $\theta=2.6$. Calculations are performed on clusters with 36, 36, 49 unit cells respectively.}
    \label{Fig:sq}
\end{figure*}

We now optimize the neural wavefunction ansatz $\Psi_{NN}(\rl)$ with parameter $\theta$ by minimizing the energy
based on the variational principle: 
\begin{equation}
    E(\theta) = \frac{\int d\bm{r}\Psi^{*}_{\theta}(\bm{r})H\Psi_{\theta}(\bm{r})}{\int d\bm{r}\Psi^{*}_{\theta}(\bm{r})\Psi_{\theta}(\bm{r})}. \label{NN_H}
\end{equation}
We use Markov Chain Monte Carlo with position and layer spin sampling to efficiently estimate the energy. The parameters $\theta$ are updated with a natural gradient method~\cite{sorella1998green}. The total optimization cost per step scale as $\mathcal{O}(N^4)$ since the the Markov chain may take $\mathcal{O}(N)$ evaluation of the neural wavefunction to equilibriate and a single evaluation cost of the many-body neural Bloch wavefunction of $N$ particles scales as $\mathcal{O}(N^3)$ due to the determinant structure.  Any local observable such as the structure factor can be computed efficiently using Monte Carlo sampling after the neural wavefunction $\Psi_{NN}(\rl)$ is obtained to probe the physics of the system. Our approach offers an efficient approach with polynomial scaling to solve many-electron problems in the twisted quantum material system. The details of the optimization is also provided in the Supplementary Materials.

{\it Results---.} We compare the ground state (GS) energies of neural Bloch wavefunction with band-projected ED. The band projection neglects band mixing between all bands, which is accurate only when the characteristic Coulomb energy $\frac{e^2}{\epsilon a_M}$ is small compared to the moiré band gap. 
More generally, band-projected ED is a variational method, and its GS energy improves with the number of bands projected onto. 

First, we compare GS energies per particle of neural wavefunction at twist $\theta=3.0$, $\epsilon=5$ and $\epsilon=10$ for $\nu=\frac{2}{3}$ on a $3 \times 3$ cluster with from 1-band to 5-band ED in Fig.~\ref{Fig3: eps}. It is clear that neural Bloch wavefunction and higher band ED outperform 1-band ED significantly, lowering the energy per particle by $\sim 6$meV for $\epsilon=5$. This indicates the presence of strong band mixing. We note that NN achieves lower energies even compared to 5-band ED, demonstrating that NN is able to capture the influence of remote bands. We also present the electron density from NN optimization. Remarkably, although we do not impose any translational symmetry on the NN, the optimization leads to translationally invariant charge density on the moir\'e superlattice.

We then compare GS energies as a function of filling $\nu$ at twist $\theta = 3.0$, and $\epsilon=5$, using a 27-unit-cell cluster ($3\sqrt{3} \times 3\sqrt{3}$) for $\nu=\frac{1}{3},\frac{5}{9},\frac{2}{3}$, 25 ($5\times 5$) for $\frac{3}{5}$ and 28 ($2\sqrt{7} \times 2\sqrt{7}$) for $\frac{4}{7}$. 
In Fig. \ref{Fig2: filling}, we show the superior performance of the neural Bloch wavefunction compared to band projected ED:  our NN energies are consistently lower by several meV per particle.

To probe the nature of the many-body ground states at various fractional fillings, 
we calculate the static structure factor with the optimized NN wavefunctions. The (full) structure factor $S(\mathbf{q})$ is generally defined as the Fourier transform of the static density-density correlation function:

\begin{equation}
    \chi (\mathbf{r}_i, \mathbf{r}_j) = \langle\rho(\mathbf{r}_i)\rho(\mathbf{r}_j)\rangle =\frac{1}{A}\sum_{\mathbf{q}}e^{i\mathbf{q}(\mathbf{r}_j-\mathbf{r}_i))}S(\mathbf{q}).
\end{equation}

 Using the neural Bloch wavefunction on 36-unit-cell cluster ($3 \times 3$) and 49-unit-cell ($7 \times 7$) cluster, we plot the structure factor for different fillings at $\theta=2.6$ and $\epsilon=10$ in Fig. \ref{Fig:sq}. 
$S(\mathbf{q})$ is plotted vs $a_M q$ along two directions, corresponding to horizontal line cuts from $\Gamma$ point to $K$ point and $M$ point respectively, which are shown in the 2d colorplot inset. Fig. \ref{Fig:sq} clearly demonstrates the quadratic dependence for small $q$, as expected for insulating states \cite{onishi2024quantum}. The data also reveals that $S(q)$ approaches to 1 at large $q$, which is the expected general behavior of full structure factor. 

For $\nu=\frac{1}{3}$ we see pronounced Bragg peaks at exactly K, K' points, demonstrating the CDW phase at this filling. In constrast, no Bragg peaks are present at $\nu=\frac{2}{3},\frac{4}{7}$. At these filling fractions,  the presence of energy gap, the absence of Bragg peak, and the resemblance of $S(q)$ with the corresponding fractional quantum Hall states are consistent with the FQAH liquid. 

We further perform polynomial fitting to extract the $q^2$ coefficient of $S(q)$ at small $q$, known as quantum weight \cite{onishi2023quantum,zaklama2024structure}:
\bea
    S(\mathbf{q})= \frac{K_{\alpha \beta}}{2\pi}q_\alpha q_\beta + \cdots .
\eea
It has been recently shown that the quantum weight $K$ of (integer or fractional) Chern insulators has a universal 
lower bound $K\geq C$ where $C$ is the many-body Chern number \cite{onishi2024StructureFact}. 
The quantum weight we extract for $\nu=\frac{2}{3},\frac{4}{7}$ are 0.667 and 0.636 respectively, which indeed satisfy the topological bound. 

Our results demonstrate the capability of the neural Bloch wavefunction to faithfully capture important physical observables of symmetry breaking electron crystals and fractional quantum liquids, in addition to  accurately attaining their ground state energies.

{\it Discussion---.} In this work, we simulate the fractional states using a many-body neural Bloch wave function, benchmarked against ED, demonstrating the neural network's superior computational performance and its ability to capture both electron crystals and fractional quantum liquids in a unified manner. Focusing on MoTe$_2$ at various fractional filling \cite{reddy2023fractional}, we show the neural Bloch wavefunction outperforms in both band mixing and system size, yielding lower variational energies and accurately calculating the structure factor and quantum weight. Neural network-based wavefunctions offer the accuracy and flexibility needed to solve strongly interacting quantum materials, offering important insights into fractional and topological systems. 

\par Looking ahead, we would like to apply our method to study fractionalized metallic states and putative non-Abelian FCIs in twisted TMDs at even-denominator filling fractions, such as $\nu=1/2$ and $3/2$. Another exciting direction is to extend our NN architecture to include spin degrees of freedom and study fractional quantum spin Hall states such as $\nu=3$ \cite{kang2024observationMoTe2}.  

{\it Acknowledgement---.} The authors acknowledge helpful discussions with Xiang Li, Yixiao Chen, Weiluo Ren, Xiuhan Hu, Changnan Peng, Zhuo Chen, Yugo Onishi, and Aidan Reddy. We acknowledge Aidan Reddy for sharing code used for the exact diagonalization calculations. This work was primarily supported by the Air Force Office of Scientific Research under award number FA2386-24-1-4043.  
DL acknowledges support from the National Science Foundation under Cooperative Agreement PHY-2019786 (The NSF AI Institute for Artificial Intelligence and Fundamental Interactions,
http://iaifi.org/). TZ was supported by the MIT Dean of Science Graduate Student Fellowship. LF was supported in part by a Simons Investigator Award from the Simons Foundation and the NSF through Award No. PHY 2425180.

\bibliography{ref}

\begin{thebibliography}{69}%
\makeatletter
\providecommand \@ifxundefined [1]{%
 \@ifx{#1\undefined}
}%
\providecommand \@ifnum [1]{%
 \ifnum #1\expandafter \@firstoftwo
 \else \expandafter \@secondoftwo
 \fi
}%
\providecommand \@ifx [1]{%
 \ifx #1\expandafter \@firstoftwo
 \else \expandafter \@secondoftwo
 \fi
}%
\providecommand \natexlab [1]{#1}%
\providecommand \enquote  [1]{``#1''}%
\providecommand \bibnamefont  [1]{#1}%
\providecommand \bibfnamefont [1]{#1}%
\providecommand \citenamefont [1]{#1}%
\providecommand \href@noop [0]{\@secondoftwo}%
\providecommand \href [0]{\begingroup \@sanitize@url \@href}%
\providecommand \@href[1]{\@@startlink{#1}\@@href}%
\providecommand \@@href[1]{\endgroup#1\@@endlink}%
\providecommand \@sanitize@url [0]{\catcode `\\12\catcode `\$12\catcode `\&12\catcode `\#12\catcode `\^12\catcode `\_12\catcode `\%12\relax}%
\providecommand \@@startlink[1]{}%
\providecommand \@@endlink[0]{}%
\providecommand \url  [0]{\begingroup\@sanitize@url \@url }%
\providecommand \@url [1]{\endgroup\@href {#1}{\urlprefix }}%
\providecommand \urlprefix  [0]{URL }%
\providecommand \Eprint [0]{\href }%
\providecommand \doibase [0]{https://doi.org/}%
\providecommand \selectlanguage [0]{\@gobble}%
\providecommand \bibinfo  [0]{\@secondoftwo}%
\providecommand \bibfield  [0]{\@secondoftwo}%
\providecommand \translation [1]{[#1]}%
\providecommand \BibitemOpen [0]{}%
\providecommand \bibitemStop [0]{}%
\providecommand \bibitemNoStop [0]{.\EOS\space}%
\providecommand \EOS [0]{\spacefactor3000\relax}%
\providecommand \BibitemShut  [1]{\csname bibitem#1\endcsname}%
\let\auto@bib@innerbib\@empty
\bibitem [{\citenamefont {Regan}\ \emph {et~al.}(2020)\citenamefont {Regan}, \citenamefont {Wang}, \citenamefont {Jin}, \citenamefont {Bakti~Utama}, \citenamefont {Gao}, \citenamefont {Wei}, \citenamefont {Zhao}, \citenamefont {Zhao}, \citenamefont {Zhang}, \citenamefont {Yumigeta} \emph {et~al.}}]{regan2020mott}%
  \BibitemOpen
  \bibfield  {author} {\bibinfo {author} {\bibfnamefont {E.~C.}\ \bibnamefont {Regan}}, \bibinfo {author} {\bibfnamefont {D.}~\bibnamefont {Wang}}, \bibinfo {author} {\bibfnamefont {C.}~\bibnamefont {Jin}}, \bibinfo {author} {\bibfnamefont {M.~I.}\ \bibnamefont {Bakti~Utama}}, \bibinfo {author} {\bibfnamefont {B.}~\bibnamefont {Gao}}, \bibinfo {author} {\bibfnamefont {X.}~\bibnamefont {Wei}}, \bibinfo {author} {\bibfnamefont {S.}~\bibnamefont {Zhao}}, \bibinfo {author} {\bibfnamefont {W.}~\bibnamefont {Zhao}}, \bibinfo {author} {\bibfnamefont {Z.}~\bibnamefont {Zhang}}, \bibinfo {author} {\bibfnamefont {K.}~\bibnamefont {Yumigeta}}, \emph {et~al.},\ }\bibfield  {title} {\bibinfo {title} {Mott and generalized wigner crystal states in wse2/ws2 moir{\'e} superlattices},\ }\href {https://doi.org/10.1038/s41586-020-2092-4} {\bibfield  {journal} {\bibinfo  {journal} {Nature}\ }\textbf {\bibinfo {volume} {579}},\ \bibinfo {pages} {359} (\bibinfo {year} {2020})}\BibitemShut {NoStop}%
\bibitem [{\citenamefont {Xia}\ \emph {et~al.}(2024)\citenamefont {Xia}, \citenamefont {Han}, \citenamefont {Watanabe}, \citenamefont {Taniguchi}, \citenamefont {Shan},\ and\ \citenamefont {Mak}}]{xia2024unconventional}%
  \BibitemOpen
  \bibfield  {author} {\bibinfo {author} {\bibfnamefont {Y.}~\bibnamefont {Xia}}, \bibinfo {author} {\bibfnamefont {Z.}~\bibnamefont {Han}}, \bibinfo {author} {\bibfnamefont {K.}~\bibnamefont {Watanabe}}, \bibinfo {author} {\bibfnamefont {T.}~\bibnamefont {Taniguchi}}, \bibinfo {author} {\bibfnamefont {J.}~\bibnamefont {Shan}},\ and\ \bibinfo {author} {\bibfnamefont {K.~F.}\ \bibnamefont {Mak}},\ }\bibfield  {title} {\bibinfo {title} {Unconventional superconductivity in twisted bilayer wse2},\ }\href@noop {} {\bibfield  {journal} {\bibinfo  {journal} {arXiv preprint arXiv:2405.14784}\ } (\bibinfo {year} {2024})}\BibitemShut {NoStop}%
\bibitem [{\citenamefont {Guo}\ \emph {et~al.}(2024)\citenamefont {Guo}, \citenamefont {Pack}, \citenamefont {Swann}, \citenamefont {Holtzman}, \citenamefont {Cothrine}, \citenamefont {Watanabe}, \citenamefont {Taniguchi}, \citenamefont {Mandrus}, \citenamefont {Barmak}, \citenamefont {Hone}, \citenamefont {Millis}, \citenamefont {Pasupathy},\ and\ \citenamefont {Dean}}]{guo2024superconductivity}%
  \BibitemOpen
  \bibfield  {author} {\bibinfo {author} {\bibfnamefont {Y.}~\bibnamefont {Guo}}, \bibinfo {author} {\bibfnamefont {J.}~\bibnamefont {Pack}}, \bibinfo {author} {\bibfnamefont {J.}~\bibnamefont {Swann}}, \bibinfo {author} {\bibfnamefont {L.}~\bibnamefont {Holtzman}}, \bibinfo {author} {\bibfnamefont {M.}~\bibnamefont {Cothrine}}, \bibinfo {author} {\bibfnamefont {K.}~\bibnamefont {Watanabe}}, \bibinfo {author} {\bibfnamefont {T.}~\bibnamefont {Taniguchi}}, \bibinfo {author} {\bibfnamefont {D.}~\bibnamefont {Mandrus}}, \bibinfo {author} {\bibfnamefont {K.}~\bibnamefont {Barmak}}, \bibinfo {author} {\bibfnamefont {J.}~\bibnamefont {Hone}}, \bibinfo {author} {\bibfnamefont {A.~J.}\ \bibnamefont {Millis}}, \bibinfo {author} {\bibfnamefont {A.~N.}\ \bibnamefont {Pasupathy}},\ and\ \bibinfo {author} {\bibfnamefont {C.~R.}\ \bibnamefont {Dean}},\ }\bibfield  {title} {\bibinfo {title} {Superconductivity in twisted bilayer wse2},\ }\href@noop {} {\bibfield  {journal} {\bibinfo  {journal} {arXiv preprint
  arXiv:2406.03418}\ } (\bibinfo {year} {2024})}\BibitemShut {NoStop}%
\bibitem [{\citenamefont {Li}\ \emph {et~al.}(2021{\natexlab{a}})\citenamefont {Li}, \citenamefont {Jiang}, \citenamefont {Shen}, \citenamefont {Zhang}, \citenamefont {Li}, \citenamefont {Tao}, \citenamefont {Devakul}, \citenamefont {Watanabe}, \citenamefont {Taniguchi}, \citenamefont {Fu} \emph {et~al.}}]{li2021quantum}%
  \BibitemOpen
  \bibfield  {author} {\bibinfo {author} {\bibfnamefont {T.}~\bibnamefont {Li}}, \bibinfo {author} {\bibfnamefont {S.}~\bibnamefont {Jiang}}, \bibinfo {author} {\bibfnamefont {B.}~\bibnamefont {Shen}}, \bibinfo {author} {\bibfnamefont {Y.}~\bibnamefont {Zhang}}, \bibinfo {author} {\bibfnamefont {L.}~\bibnamefont {Li}}, \bibinfo {author} {\bibfnamefont {Z.}~\bibnamefont {Tao}}, \bibinfo {author} {\bibfnamefont {T.}~\bibnamefont {Devakul}}, \bibinfo {author} {\bibfnamefont {K.}~\bibnamefont {Watanabe}}, \bibinfo {author} {\bibfnamefont {T.}~\bibnamefont {Taniguchi}}, \bibinfo {author} {\bibfnamefont {L.}~\bibnamefont {Fu}}, \emph {et~al.},\ }\bibfield  {title} {\bibinfo {title} {Quantum anomalous hall effect from intertwined moir{\'e} bands},\ }\href {https://doi.org/10.1038/s41586-021-04171-1} {\bibfield  {journal} {\bibinfo  {journal} {Nature}\ }\textbf {\bibinfo {volume} {600}},\ \bibinfo {pages} {641} (\bibinfo {year} {2021}{\natexlab{a}})}\BibitemShut {NoStop}%
\bibitem [{\citenamefont {Cai}\ \emph {et~al.}(2023)\citenamefont {Cai}, \citenamefont {Anderson}, \citenamefont {Wang}, \citenamefont {Zhang}, \citenamefont {Liu}, \citenamefont {Holtzmann}, \citenamefont {Zhang}, \citenamefont {Fan}, \citenamefont {Taniguchi}, \citenamefont {Watanabe} \emph {et~al.}}]{cai2023signatures}%
  \BibitemOpen
  \bibfield  {author} {\bibinfo {author} {\bibfnamefont {J.}~\bibnamefont {Cai}}, \bibinfo {author} {\bibfnamefont {E.}~\bibnamefont {Anderson}}, \bibinfo {author} {\bibfnamefont {C.}~\bibnamefont {Wang}}, \bibinfo {author} {\bibfnamefont {X.}~\bibnamefont {Zhang}}, \bibinfo {author} {\bibfnamefont {X.}~\bibnamefont {Liu}}, \bibinfo {author} {\bibfnamefont {W.}~\bibnamefont {Holtzmann}}, \bibinfo {author} {\bibfnamefont {Y.}~\bibnamefont {Zhang}}, \bibinfo {author} {\bibfnamefont {F.}~\bibnamefont {Fan}}, \bibinfo {author} {\bibfnamefont {T.}~\bibnamefont {Taniguchi}}, \bibinfo {author} {\bibfnamefont {K.}~\bibnamefont {Watanabe}}, \emph {et~al.},\ }\bibfield  {title} {\bibinfo {title} {Signatures of fractional quantum anomalous hall states in twisted mote2},\ }\href {https://doi.org/10.1038/s41586-023-06289-w} {\bibfield  {journal} {\bibinfo  {journal} {Nature}\ ,\ \bibinfo {pages} {1}} (\bibinfo {year} {2023})}\BibitemShut {NoStop}%
\bibitem [{\citenamefont {Park}\ \emph {et~al.}(2023)\citenamefont {Park}, \citenamefont {Cai}, \citenamefont {Anderson}, \citenamefont {Zhang}, \citenamefont {Zhu}, \citenamefont {Liu}, \citenamefont {Wang}, \citenamefont {Holtzmann}, \citenamefont {Hu}, \citenamefont {Liu} \emph {et~al.}}]{park2023observation}%
  \BibitemOpen
  \bibfield  {author} {\bibinfo {author} {\bibfnamefont {H.}~\bibnamefont {Park}}, \bibinfo {author} {\bibfnamefont {J.}~\bibnamefont {Cai}}, \bibinfo {author} {\bibfnamefont {E.}~\bibnamefont {Anderson}}, \bibinfo {author} {\bibfnamefont {Y.}~\bibnamefont {Zhang}}, \bibinfo {author} {\bibfnamefont {J.}~\bibnamefont {Zhu}}, \bibinfo {author} {\bibfnamefont {X.}~\bibnamefont {Liu}}, \bibinfo {author} {\bibfnamefont {C.}~\bibnamefont {Wang}}, \bibinfo {author} {\bibfnamefont {W.}~\bibnamefont {Holtzmann}}, \bibinfo {author} {\bibfnamefont {C.}~\bibnamefont {Hu}}, \bibinfo {author} {\bibfnamefont {Z.}~\bibnamefont {Liu}}, \emph {et~al.},\ }\bibfield  {title} {\bibinfo {title} {Observation of fractionally quantized anomalous hall effect},\ }\href {https://doi.org/10.1038/s41586-023-06536-0} {\bibfield  {journal} {\bibinfo  {journal} {Nature}\ ,\ \bibinfo {pages} {1}} (\bibinfo {year} {2023})}\BibitemShut {NoStop}%
\bibitem [{\citenamefont {Zeng}\ \emph {et~al.}(2023)\citenamefont {Zeng}, \citenamefont {Xia}, \citenamefont {Kang}, \citenamefont {Zhu}, \citenamefont {Kn{\"u}ppel}, \citenamefont {Vaswani}, \citenamefont {Watanabe}, \citenamefont {Taniguchi}, \citenamefont {Mak},\ and\ \citenamefont {Shan}}]{zeng2023thermodynamic}%
  \BibitemOpen
  \bibfield  {author} {\bibinfo {author} {\bibfnamefont {Y.}~\bibnamefont {Zeng}}, \bibinfo {author} {\bibfnamefont {Z.}~\bibnamefont {Xia}}, \bibinfo {author} {\bibfnamefont {K.}~\bibnamefont {Kang}}, \bibinfo {author} {\bibfnamefont {J.}~\bibnamefont {Zhu}}, \bibinfo {author} {\bibfnamefont {P.}~\bibnamefont {Kn{\"u}ppel}}, \bibinfo {author} {\bibfnamefont {C.}~\bibnamefont {Vaswani}}, \bibinfo {author} {\bibfnamefont {K.}~\bibnamefont {Watanabe}}, \bibinfo {author} {\bibfnamefont {T.}~\bibnamefont {Taniguchi}}, \bibinfo {author} {\bibfnamefont {K.~F.}\ \bibnamefont {Mak}},\ and\ \bibinfo {author} {\bibfnamefont {J.}~\bibnamefont {Shan}},\ }\bibfield  {title} {\bibinfo {title} {Thermodynamic evidence of fractional chern insulator in moir{\'e} mote2},\ }\href {https://www.nature.com/articles/s41586-023-06452-3} {\bibfield  {journal} {\bibinfo  {journal} {Nature}\ ,\ \bibinfo {pages} {1}} (\bibinfo {year} {2023})}\BibitemShut {NoStop}%
\bibitem [{\citenamefont {Xu}\ \emph {et~al.}(2023)\citenamefont {Xu}, \citenamefont {Sun}, \citenamefont {Jia}, \citenamefont {Liu}, \citenamefont {Xu}, \citenamefont {Li}, \citenamefont {Gu}, \citenamefont {Watanabe}, \citenamefont {Taniguchi}, \citenamefont {Tong} \emph {et~al.}}]{xu2023observation}%
  \BibitemOpen
  \bibfield  {author} {\bibinfo {author} {\bibfnamefont {F.}~\bibnamefont {Xu}}, \bibinfo {author} {\bibfnamefont {Z.}~\bibnamefont {Sun}}, \bibinfo {author} {\bibfnamefont {T.}~\bibnamefont {Jia}}, \bibinfo {author} {\bibfnamefont {C.}~\bibnamefont {Liu}}, \bibinfo {author} {\bibfnamefont {C.}~\bibnamefont {Xu}}, \bibinfo {author} {\bibfnamefont {C.}~\bibnamefont {Li}}, \bibinfo {author} {\bibfnamefont {Y.}~\bibnamefont {Gu}}, \bibinfo {author} {\bibfnamefont {K.}~\bibnamefont {Watanabe}}, \bibinfo {author} {\bibfnamefont {T.}~\bibnamefont {Taniguchi}}, \bibinfo {author} {\bibfnamefont {B.}~\bibnamefont {Tong}}, \emph {et~al.},\ }\bibfield  {title} {\bibinfo {title} {Observation of integer and fractional quantum anomalous hall states in twisted bilayer mote2},\ }\href@noop {} {\bibfield  {journal} {\bibinfo  {journal} {arXiv preprint arXiv:2308.06177}\ } (\bibinfo {year} {2023})}\BibitemShut {NoStop}%
\bibitem [{\citenamefont {Morales-Dur{\'a}n}\ \emph {et~al.}(2021)\citenamefont {Morales-Dur{\'a}n}, \citenamefont {MacDonald},\ and\ \citenamefont {Potasz}}]{morales2021metal}%
  \BibitemOpen
  \bibfield  {author} {\bibinfo {author} {\bibfnamefont {N.}~\bibnamefont {Morales-Dur{\'a}n}}, \bibinfo {author} {\bibfnamefont {A.~H.}\ \bibnamefont {MacDonald}},\ and\ \bibinfo {author} {\bibfnamefont {P.}~\bibnamefont {Potasz}},\ }\bibfield  {title} {\bibinfo {title} {Metal-insulator transition in transition metal dichalcogenide heterobilayer moir{\'e} superlattices},\ }\href@noop {} {\bibfield  {journal} {\bibinfo  {journal} {Physical Review B}\ }\textbf {\bibinfo {volume} {103}},\ \bibinfo {pages} {L241110} (\bibinfo {year} {2021})}\BibitemShut {NoStop}%
\bibitem [{\citenamefont {Li}\ \emph {et~al.}(2021{\natexlab{b}})\citenamefont {Li}, \citenamefont {Kumar}, \citenamefont {Sun},\ and\ \citenamefont {Lin}}]{li2021spontaneous}%
  \BibitemOpen
  \bibfield  {author} {\bibinfo {author} {\bibfnamefont {H.}~\bibnamefont {Li}}, \bibinfo {author} {\bibfnamefont {U.}~\bibnamefont {Kumar}}, \bibinfo {author} {\bibfnamefont {K.}~\bibnamefont {Sun}},\ and\ \bibinfo {author} {\bibfnamefont {S.-Z.}\ \bibnamefont {Lin}},\ }\bibfield  {title} {\bibinfo {title} {Spontaneous fractional chern insulators in transition metal dichalcogenide moir{\'e} superlattices},\ }\href {https://doi.org/10.1103/PhysRevResearch.3.L032070} {\bibfield  {journal} {\bibinfo  {journal} {Physical Review Research}\ }\textbf {\bibinfo {volume} {3}},\ \bibinfo {pages} {L032070} (\bibinfo {year} {2021}{\natexlab{b}})}\BibitemShut {NoStop}%
\bibitem [{\citenamefont {Cr{\'e}pel}\ and\ \citenamefont {Fu}(2023)}]{crepel2023anomalous}%
  \BibitemOpen
  \bibfield  {author} {\bibinfo {author} {\bibfnamefont {V.}~\bibnamefont {Cr{\'e}pel}}\ and\ \bibinfo {author} {\bibfnamefont {L.}~\bibnamefont {Fu}},\ }\bibfield  {title} {\bibinfo {title} {Anomalous hall metal and fractional chern insulator in twisted transition metal dichalcogenides},\ }\href {https://doi.org/10.1103/PhysRevB.107.L201109} {\bibfield  {journal} {\bibinfo  {journal} {Physical Review B}\ }\textbf {\bibinfo {volume} {107}},\ \bibinfo {pages} {L201109} (\bibinfo {year} {2023})}\BibitemShut {NoStop}%
\bibitem [{\citenamefont {Reddy}\ \emph {et~al.}(2023{\natexlab{a}})\citenamefont {Reddy}, \citenamefont {Alsallom}, \citenamefont {Zhang}, \citenamefont {Devakul},\ and\ \citenamefont {Fu}}]{reddy2023fractional}%
  \BibitemOpen
  \bibfield  {author} {\bibinfo {author} {\bibfnamefont {A.~P.}\ \bibnamefont {Reddy}}, \bibinfo {author} {\bibfnamefont {F.}~\bibnamefont {Alsallom}}, \bibinfo {author} {\bibfnamefont {Y.}~\bibnamefont {Zhang}}, \bibinfo {author} {\bibfnamefont {T.}~\bibnamefont {Devakul}},\ and\ \bibinfo {author} {\bibfnamefont {L.}~\bibnamefont {Fu}},\ }\bibfield  {title} {\bibinfo {title} {Fractional quantum anomalous hall states in twisted bilayer ${\mathrm{mote}}_{2}$ and ${\mathrm{wse}}_{2}$},\ }\href {https://doi.org/10.1103/PhysRevB.108.085117} {\bibfield  {journal} {\bibinfo  {journal} {Phys. Rev. B}\ }\textbf {\bibinfo {volume} {108}},\ \bibinfo {pages} {085117} (\bibinfo {year} {2023}{\natexlab{a}})}\BibitemShut {NoStop}%
\bibitem [{\citenamefont {Wang}\ \emph {et~al.}(2023)\citenamefont {Wang}, \citenamefont {Zhang}, \citenamefont {Liu}, \citenamefont {He}, \citenamefont {Xu}, \citenamefont {Ran}, \citenamefont {Cao},\ and\ \citenamefont {Xiao}}]{wang2023fractional}%
  \BibitemOpen
  \bibfield  {author} {\bibinfo {author} {\bibfnamefont {C.}~\bibnamefont {Wang}}, \bibinfo {author} {\bibfnamefont {X.-W.}\ \bibnamefont {Zhang}}, \bibinfo {author} {\bibfnamefont {X.}~\bibnamefont {Liu}}, \bibinfo {author} {\bibfnamefont {Y.}~\bibnamefont {He}}, \bibinfo {author} {\bibfnamefont {X.}~\bibnamefont {Xu}}, \bibinfo {author} {\bibfnamefont {Y.}~\bibnamefont {Ran}}, \bibinfo {author} {\bibfnamefont {T.}~\bibnamefont {Cao}},\ and\ \bibinfo {author} {\bibfnamefont {D.}~\bibnamefont {Xiao}},\ }\bibfield  {title} {\bibinfo {title} {Fractional chern insulator in twisted bilayer mote $ \_2$},\ }\href {https://arxiv.org/abs/2304.11864} {\bibfield  {journal} {\bibinfo  {journal} {arXiv preprint arXiv:2304.11864}\ } (\bibinfo {year} {2023})}\BibitemShut {NoStop}%
\bibitem [{\citenamefont {Zhang}\ \emph {et~al.}(2020)\citenamefont {Zhang}, \citenamefont {Yuan},\ and\ \citenamefont {Fu}}]{zhang2020moire}%
  \BibitemOpen
  \bibfield  {author} {\bibinfo {author} {\bibfnamefont {Y.}~\bibnamefont {Zhang}}, \bibinfo {author} {\bibfnamefont {N.~F.}\ \bibnamefont {Yuan}},\ and\ \bibinfo {author} {\bibfnamefont {L.}~\bibnamefont {Fu}},\ }\bibfield  {title} {\bibinfo {title} {Moir{\'e} quantum chemistry: charge transfer in transition metal dichalcogenide superlattices},\ }\href@noop {} {\bibfield  {journal} {\bibinfo  {journal} {Physical Review B}\ }\textbf {\bibinfo {volume} {102}},\ \bibinfo {pages} {201115} (\bibinfo {year} {2020})}\BibitemShut {NoStop}%
\bibitem [{\citenamefont {Reddy}\ \emph {et~al.}(2023{\natexlab{b}})\citenamefont {Reddy}, \citenamefont {Devakul},\ and\ \citenamefont {Fu}}]{reddy2023artificial}%
  \BibitemOpen
  \bibfield  {author} {\bibinfo {author} {\bibfnamefont {A.~P.}\ \bibnamefont {Reddy}}, \bibinfo {author} {\bibfnamefont {T.}~\bibnamefont {Devakul}},\ and\ \bibinfo {author} {\bibfnamefont {L.}~\bibnamefont {Fu}},\ }\bibfield  {title} {\bibinfo {title} {Artificial atoms, wigner molecules, and emergent kagome lattice in semiconductor moir´e superlattices},\ }\href@noop {} {\bibfield  {journal} {\bibinfo  {journal} {arXiv preprint arXiv:2301.00799}\ } (\bibinfo {year} {2023}{\natexlab{b}})}\BibitemShut {NoStop}%
\bibitem [{\citenamefont {Abouelkomsan}\ \emph {et~al.}(2024)\citenamefont {Abouelkomsan}, \citenamefont {Reddy}, \citenamefont {Fu},\ and\ \citenamefont {Bergholtz}}]{abouelkomsan2024band}%
  \BibitemOpen
  \bibfield  {author} {\bibinfo {author} {\bibfnamefont {A.}~\bibnamefont {Abouelkomsan}}, \bibinfo {author} {\bibfnamefont {A.~P.}\ \bibnamefont {Reddy}}, \bibinfo {author} {\bibfnamefont {L.}~\bibnamefont {Fu}},\ and\ \bibinfo {author} {\bibfnamefont {E.~J.}\ \bibnamefont {Bergholtz}},\ }\bibfield  {title} {\bibinfo {title} {Band mixing in the quantum anomalous hall regime of twisted semiconductor bilayers},\ }\href@noop {} {\bibfield  {journal} {\bibinfo  {journal} {Physical Review B}\ }\textbf {\bibinfo {volume} {109}},\ \bibinfo {pages} {L121107} (\bibinfo {year} {2024})}\BibitemShut {NoStop}%
\bibitem [{\citenamefont {Yu}\ \emph {et~al.}(2024{\natexlab{a}})\citenamefont {Yu}, \citenamefont {Herzog-Arbeitman}, \citenamefont {Wang}, \citenamefont {Vafek}, \citenamefont {Bernevig},\ and\ \citenamefont {Regnault}}]{yu2024fractional}%
  \BibitemOpen
  \bibfield  {author} {\bibinfo {author} {\bibfnamefont {J.}~\bibnamefont {Yu}}, \bibinfo {author} {\bibfnamefont {J.}~\bibnamefont {Herzog-Arbeitman}}, \bibinfo {author} {\bibfnamefont {M.}~\bibnamefont {Wang}}, \bibinfo {author} {\bibfnamefont {O.}~\bibnamefont {Vafek}}, \bibinfo {author} {\bibfnamefont {B.~A.}\ \bibnamefont {Bernevig}},\ and\ \bibinfo {author} {\bibfnamefont {N.}~\bibnamefont {Regnault}},\ }\bibfield  {title} {\bibinfo {title} {Fractional chern insulators versus nonmagnetic states in twisted bilayer mote 2},\ }\href@noop {} {\bibfield  {journal} {\bibinfo  {journal} {Physical Review B}\ }\textbf {\bibinfo {volume} {109}},\ \bibinfo {pages} {045147} (\bibinfo {year} {2024}{\natexlab{a}})}\BibitemShut {NoStop}%
\bibitem [{\citenamefont {Carleo}\ and\ \citenamefont {Troyer}(2017{\natexlab{a}})}]{Carleo602}%
  \BibitemOpen
  \bibfield  {author} {\bibinfo {author} {\bibfnamefont {G.}~\bibnamefont {Carleo}}\ and\ \bibinfo {author} {\bibfnamefont {M.}~\bibnamefont {Troyer}},\ }\bibfield  {title} {\bibinfo {title} {Solving the quantum many-body problem with artificial neural networks},\ }\href {https://doi.org/10.1126/science.aag2302} {\bibfield  {journal} {\bibinfo  {journal} {Science}\ }\textbf {\bibinfo {volume} {355}},\ \bibinfo {pages} {602} (\bibinfo {year} {2017}{\natexlab{a}})},\ \Eprint {https://arxiv.org/abs/https://www.science.org/doi/pdf/10.1126/science.aag2302} {https://www.science.org/doi/pdf/10.1126/science.aag2302} \BibitemShut {NoStop}%
\bibitem [{\citenamefont {Chen}\ \emph {et~al.}(2023)\citenamefont {Chen}, \citenamefont {Newhouse}, \citenamefont {Chen}, \citenamefont {Luo},\ and\ \citenamefont {Soljacic}}]{chen2023autoregressive}%
  \BibitemOpen
  \bibfield  {author} {\bibinfo {author} {\bibfnamefont {Z.}~\bibnamefont {Chen}}, \bibinfo {author} {\bibfnamefont {L.}~\bibnamefont {Newhouse}}, \bibinfo {author} {\bibfnamefont {E.}~\bibnamefont {Chen}}, \bibinfo {author} {\bibfnamefont {D.}~\bibnamefont {Luo}},\ and\ \bibinfo {author} {\bibfnamefont {M.}~\bibnamefont {Soljacic}},\ }\bibfield  {title} {\bibinfo {title} {Autoregressive neural tensornet: Bridging neural networks and tensor networks for quantum many-body simulation},\ }\href@noop {} {\bibfield  {journal} {\bibinfo  {journal} {arXiv preprint arXiv:2304.01996}\ } (\bibinfo {year} {2023})}\BibitemShut {NoStop}%
\bibitem [{\citenamefont {Robledo~Moreno}\ \emph {et~al.}(2022)\citenamefont {Robledo~Moreno}, \citenamefont {Carleo}, \citenamefont {Georges},\ and\ \citenamefont {Stokes}}]{robledo2022fermionic}%
  \BibitemOpen
  \bibfield  {author} {\bibinfo {author} {\bibfnamefont {J.}~\bibnamefont {Robledo~Moreno}}, \bibinfo {author} {\bibfnamefont {G.}~\bibnamefont {Carleo}}, \bibinfo {author} {\bibfnamefont {A.}~\bibnamefont {Georges}},\ and\ \bibinfo {author} {\bibfnamefont {J.}~\bibnamefont {Stokes}},\ }\bibfield  {title} {\bibinfo {title} {Fermionic wave functions from neural-network constrained hidden states},\ }\href@noop {} {\bibfield  {journal} {\bibinfo  {journal} {Proceedings of the National Academy of Sciences}\ }\textbf {\bibinfo {volume} {119}},\ \bibinfo {pages} {e2122059119} (\bibinfo {year} {2022})}\BibitemShut {NoStop}%
\bibitem [{\citenamefont {Chen}\ \emph {et~al.}(2022)\citenamefont {Chen}, \citenamefont {Luo}, \citenamefont {Hu},\ and\ \citenamefont {Clark}}]{chen2022simulating}%
  \BibitemOpen
  \bibfield  {author} {\bibinfo {author} {\bibfnamefont {Z.}~\bibnamefont {Chen}}, \bibinfo {author} {\bibfnamefont {D.}~\bibnamefont {Luo}}, \bibinfo {author} {\bibfnamefont {K.}~\bibnamefont {Hu}},\ and\ \bibinfo {author} {\bibfnamefont {B.~K.}\ \bibnamefont {Clark}},\ }\bibfield  {title} {\bibinfo {title} {Simulating 2+ 1d lattice quantum electrodynamics at finite density with neural flow wavefunctions},\ }\href@noop {} {\bibfield  {journal} {\bibinfo  {journal} {arXiv preprint arXiv:2212.06835}\ } (\bibinfo {year} {2022})}\BibitemShut {NoStop}%
\bibitem [{\citenamefont {Carleo}\ and\ \citenamefont {Troyer}(2017{\natexlab{b}})}]{doi:10.1126/science.aag2302}%
  \BibitemOpen
  \bibfield  {author} {\bibinfo {author} {\bibfnamefont {G.}~\bibnamefont {Carleo}}\ and\ \bibinfo {author} {\bibfnamefont {M.}~\bibnamefont {Troyer}},\ }\bibfield  {title} {\bibinfo {title} {Solving the quantum many-body problem with artificial neural networks},\ }\href {https://doi.org/10.1126/science.aag2302} {\bibfield  {journal} {\bibinfo  {journal} {Science}\ }\textbf {\bibinfo {volume} {355}},\ \bibinfo {pages} {602} (\bibinfo {year} {2017}{\natexlab{b}})},\ \Eprint {https://arxiv.org/abs/https://www.science.org/doi/pdf/10.1126/science.aag2302} {https://www.science.org/doi/pdf/10.1126/science.aag2302} \BibitemShut {NoStop}%
\bibitem [{\citenamefont {Hibat-Allah}\ \emph {et~al.}(2020{\natexlab{a}})\citenamefont {Hibat-Allah}, \citenamefont {Ganahl}, \citenamefont {Hayward}, \citenamefont {Melko},\ and\ \citenamefont {Carrasquilla}}]{Hibat_Allah_2020}%
  \BibitemOpen
  \bibfield  {author} {\bibinfo {author} {\bibfnamefont {M.}~\bibnamefont {Hibat-Allah}}, \bibinfo {author} {\bibfnamefont {M.}~\bibnamefont {Ganahl}}, \bibinfo {author} {\bibfnamefont {L.~E.}\ \bibnamefont {Hayward}}, \bibinfo {author} {\bibfnamefont {R.~G.}\ \bibnamefont {Melko}},\ and\ \bibinfo {author} {\bibfnamefont {J.}~\bibnamefont {Carrasquilla}},\ }\bibfield  {title} {\bibinfo {title} {Recurrent neural network wave functions},\ }\bibfield  {journal} {\bibinfo  {journal} {Physical Review Research}\ }\textbf {\bibinfo {volume} {2}},\ \href {https://doi.org/10.1103/physrevresearch.2.023358} {10.1103/physrevresearch.2.023358} (\bibinfo {year} {2020}{\natexlab{a}})\BibitemShut {NoStop}%
\bibitem [{\citenamefont {Sharir}\ \emph {et~al.}(2020)\citenamefont {Sharir}, \citenamefont {Levine}, \citenamefont {Wies}, \citenamefont {Carleo},\ and\ \citenamefont {Shashua}}]{PhysRevLett.124.020503}%
  \BibitemOpen
  \bibfield  {author} {\bibinfo {author} {\bibfnamefont {O.}~\bibnamefont {Sharir}}, \bibinfo {author} {\bibfnamefont {Y.}~\bibnamefont {Levine}}, \bibinfo {author} {\bibfnamefont {N.}~\bibnamefont {Wies}}, \bibinfo {author} {\bibfnamefont {G.}~\bibnamefont {Carleo}},\ and\ \bibinfo {author} {\bibfnamefont {A.}~\bibnamefont {Shashua}},\ }\bibfield  {title} {\bibinfo {title} {Deep autoregressive models for the efficient variational simulation of many-body quantum systems},\ }\href {https://doi.org/10.1103/PhysRevLett.124.020503} {\bibfield  {journal} {\bibinfo  {journal} {Phys. Rev. Lett.}\ }\textbf {\bibinfo {volume} {124}},\ \bibinfo {pages} {020503} (\bibinfo {year} {2020})}\BibitemShut {NoStop}%
\bibitem [{\citenamefont {Irikura}\ and\ \citenamefont {Saito}(2020)}]{Irikura_2020}%
  \BibitemOpen
  \bibfield  {author} {\bibinfo {author} {\bibfnamefont {N.}~\bibnamefont {Irikura}}\ and\ \bibinfo {author} {\bibfnamefont {H.}~\bibnamefont {Saito}},\ }\bibfield  {title} {\bibinfo {title} {Neural-network quantum states at finite temperature},\ }\bibfield  {journal} {\bibinfo  {journal} {Physical Review Research}\ }\textbf {\bibinfo {volume} {2}},\ \href {https://doi.org/10.1103/physrevresearch.2.013284} {10.1103/physrevresearch.2.013284} (\bibinfo {year} {2020})\BibitemShut {NoStop}%
\bibitem [{\citenamefont {Lee}\ \emph {et~al.}(2021)\citenamefont {Lee}, \citenamefont {Patil}, \citenamefont {Zhang},\ and\ \citenamefont {Hsieh}}]{PhysRevResearch.3.023095}%
  \BibitemOpen
  \bibfield  {author} {\bibinfo {author} {\bibfnamefont {C.~K.}\ \bibnamefont {Lee}}, \bibinfo {author} {\bibfnamefont {P.}~\bibnamefont {Patil}}, \bibinfo {author} {\bibfnamefont {S.}~\bibnamefont {Zhang}},\ and\ \bibinfo {author} {\bibfnamefont {C.~Y.}\ \bibnamefont {Hsieh}},\ }\bibfield  {title} {\bibinfo {title} {Neural-network variational quantum algorithm for simulating many-body dynamics},\ }\href {https://doi.org/10.1103/PhysRevResearch.3.023095} {\bibfield  {journal} {\bibinfo  {journal} {Phys. Rev. Research}\ }\textbf {\bibinfo {volume} {3}},\ \bibinfo {pages} {023095} (\bibinfo {year} {2021})}\BibitemShut {NoStop}%
\bibitem [{\citenamefont {Han}\ and\ \citenamefont {Hartnoll}(2020)}]{Han_2020}%
  \BibitemOpen
  \bibfield  {author} {\bibinfo {author} {\bibfnamefont {X.}~\bibnamefont {Han}}\ and\ \bibinfo {author} {\bibfnamefont {S.~A.}\ \bibnamefont {Hartnoll}},\ }\bibfield  {title} {\bibinfo {title} {Deep quantum geometry of matrices},\ }\bibfield  {journal} {\bibinfo  {journal} {Physical Review X}\ }\textbf {\bibinfo {volume} {10}},\ \href {https://doi.org/10.1103/physrevx.10.011069} {10.1103/physrevx.10.011069} (\bibinfo {year} {2020})\BibitemShut {NoStop}%
\bibitem [{\citenamefont {Pfau}\ \emph {et~al.}(2020)\citenamefont {Pfau}, \citenamefont {Spencer}, \citenamefont {Matthews},\ and\ \citenamefont {Foulkes}}]{ferminet}%
  \BibitemOpen
  \bibfield  {author} {\bibinfo {author} {\bibfnamefont {D.}~\bibnamefont {Pfau}}, \bibinfo {author} {\bibfnamefont {J.~S.}\ \bibnamefont {Spencer}}, \bibinfo {author} {\bibfnamefont {A.~G. D.~G.}\ \bibnamefont {Matthews}},\ and\ \bibinfo {author} {\bibfnamefont {W.~M.~C.}\ \bibnamefont {Foulkes}},\ }\bibfield  {title} {\bibinfo {title} {Ab initio solution of the many-electron schr\"odinger equation with deep neural networks},\ }\href {https://doi.org/10.1103/PhysRevResearch.2.033429} {\bibfield  {journal} {\bibinfo  {journal} {Phys. Rev. Research}\ }\textbf {\bibinfo {volume} {2}},\ \bibinfo {pages} {033429} (\bibinfo {year} {2020})}\BibitemShut {NoStop}%
\bibitem [{\citenamefont {Choo}\ \emph {et~al.}(2019)\citenamefont {Choo}, \citenamefont {Neupert},\ and\ \citenamefont {Carleo}}]{Choo_2019}%
  \BibitemOpen
  \bibfield  {author} {\bibinfo {author} {\bibfnamefont {K.}~\bibnamefont {Choo}}, \bibinfo {author} {\bibfnamefont {T.}~\bibnamefont {Neupert}},\ and\ \bibinfo {author} {\bibfnamefont {G.}~\bibnamefont {Carleo}},\ }\bibfield  {title} {\bibinfo {title} {Two-dimensional frustrated j 1- j 2 model studied with neural network quantum states},\ }\href@noop {} {\bibfield  {journal} {\bibinfo  {journal} {Physical Review B}\ }\textbf {\bibinfo {volume} {100}},\ \bibinfo {pages} {125124} (\bibinfo {year} {2019})}\BibitemShut {NoStop}%
\bibitem [{\citenamefont {Hibat-Allah}\ \emph {et~al.}(2020{\natexlab{b}})\citenamefont {Hibat-Allah}, \citenamefont {Ganahl}, \citenamefont {Hayward}, \citenamefont {Melko},\ and\ \citenamefont {Carrasquilla}}]{rnn_wavefunction}%
  \BibitemOpen
  \bibfield  {author} {\bibinfo {author} {\bibfnamefont {M.}~\bibnamefont {Hibat-Allah}}, \bibinfo {author} {\bibfnamefont {M.}~\bibnamefont {Ganahl}}, \bibinfo {author} {\bibfnamefont {L.~E.}\ \bibnamefont {Hayward}}, \bibinfo {author} {\bibfnamefont {R.~G.}\ \bibnamefont {Melko}},\ and\ \bibinfo {author} {\bibfnamefont {J.}~\bibnamefont {Carrasquilla}},\ }\bibfield  {title} {\bibinfo {title} {Recurrent neural network wave functions},\ }\href {https://doi.org/10.1103/PhysRevResearch.2.023358} {\bibfield  {journal} {\bibinfo  {journal} {Phys. Rev. Research}\ }\textbf {\bibinfo {volume} {2}},\ \bibinfo {pages} {023358} (\bibinfo {year} {2020}{\natexlab{b}})}\BibitemShut {NoStop}%
\bibitem [{\citenamefont {Hermann}\ \emph {et~al.}(2019)\citenamefont {Hermann}, \citenamefont {Schätzle},\ and\ \citenamefont {Noé}}]{paulinet}%
  \BibitemOpen
  \bibfield  {author} {\bibinfo {author} {\bibfnamefont {J.}~\bibnamefont {Hermann}}, \bibinfo {author} {\bibfnamefont {Z.}~\bibnamefont {Schätzle}},\ and\ \bibinfo {author} {\bibfnamefont {F.}~\bibnamefont {Noé}},\ }\href@noop {} {\bibinfo {title} {Deep neural network solution of the electronic schrödinger equation}} (\bibinfo {year} {2019}),\ \Eprint {https://arxiv.org/abs/1909.08423} {arXiv:1909.08423 [physics.comp-ph]} \BibitemShut {NoStop}%
\bibitem [{\citenamefont {Glasser}\ \emph {et~al.}(2018{\natexlab{a}})\citenamefont {Glasser}, \citenamefont {Pancotti}, \citenamefont {August}, \citenamefont {Rodriguez},\ and\ \citenamefont {Cirac}}]{Glasser_2018}%
  \BibitemOpen
  \bibfield  {author} {\bibinfo {author} {\bibfnamefont {I.}~\bibnamefont {Glasser}}, \bibinfo {author} {\bibfnamefont {N.}~\bibnamefont {Pancotti}}, \bibinfo {author} {\bibfnamefont {M.}~\bibnamefont {August}}, \bibinfo {author} {\bibfnamefont {I.~D.}\ \bibnamefont {Rodriguez}},\ and\ \bibinfo {author} {\bibfnamefont {J.~I.}\ \bibnamefont {Cirac}},\ }\bibfield  {title} {\bibinfo {title} {Neural-network quantum states, string-bond states, and chiral topological states},\ }\bibfield  {journal} {\bibinfo  {journal} {Physical Review X}\ }\textbf {\bibinfo {volume} {8}},\ \href {https://doi.org/10.1103/physrevx.8.011006} {10.1103/physrevx.8.011006} (\bibinfo {year} {2018}{\natexlab{a}})\BibitemShut {NoStop}%
\bibitem [{\citenamefont {Stokes}\ \emph {et~al.}(2020)\citenamefont {Stokes}, \citenamefont {Moreno}, \citenamefont {Pnevmatikakis},\ and\ \citenamefont {Carleo}}]{Stokes_2020}%
  \BibitemOpen
  \bibfield  {author} {\bibinfo {author} {\bibfnamefont {J.}~\bibnamefont {Stokes}}, \bibinfo {author} {\bibfnamefont {J.~R.}\ \bibnamefont {Moreno}}, \bibinfo {author} {\bibfnamefont {E.~A.}\ \bibnamefont {Pnevmatikakis}},\ and\ \bibinfo {author} {\bibfnamefont {G.}~\bibnamefont {Carleo}},\ }\bibfield  {title} {\bibinfo {title} {Phases of two-dimensional spinless lattice fermions with first-quantized deep neural-network quantum states},\ }\bibfield  {journal} {\bibinfo  {journal} {Physical Review B}\ }\textbf {\bibinfo {volume} {102}},\ \href {https://doi.org/10.1103/physrevb.102.205122} {10.1103/physrevb.102.205122} (\bibinfo {year} {2020})\BibitemShut {NoStop}%
\bibitem [{\citenamefont {Nomura}\ \emph {et~al.}(2017)\citenamefont {Nomura}, \citenamefont {Darmawan}, \citenamefont {Yamaji},\ and\ \citenamefont {Imada}}]{Nomura_2017}%
  \BibitemOpen
  \bibfield  {author} {\bibinfo {author} {\bibfnamefont {Y.}~\bibnamefont {Nomura}}, \bibinfo {author} {\bibfnamefont {A.~S.}\ \bibnamefont {Darmawan}}, \bibinfo {author} {\bibfnamefont {Y.}~\bibnamefont {Yamaji}},\ and\ \bibinfo {author} {\bibfnamefont {M.}~\bibnamefont {Imada}},\ }\bibfield  {title} {\bibinfo {title} {Restricted boltzmann machine learning for solving strongly correlated quantum systems},\ }\bibfield  {journal} {\bibinfo  {journal} {Physical Review B}\ }\textbf {\bibinfo {volume} {96}},\ \href {https://doi.org/10.1103/physrevb.96.205152} {10.1103/physrevb.96.205152} (\bibinfo {year} {2017})\BibitemShut {NoStop}%
\bibitem [{\citenamefont {Martyn}\ \emph {et~al.}(2022)\citenamefont {Martyn}, \citenamefont {Najafi},\ and\ \citenamefont {Luo}}]{martyn2022variational}%
  \BibitemOpen
  \bibfield  {author} {\bibinfo {author} {\bibfnamefont {J.~M.}\ \bibnamefont {Martyn}}, \bibinfo {author} {\bibfnamefont {K.}~\bibnamefont {Najafi}},\ and\ \bibinfo {author} {\bibfnamefont {D.}~\bibnamefont {Luo}},\ }\bibfield  {title} {\bibinfo {title} {Variational neural-network ansatz for continuum quantum field theory},\ }\href@noop {} {\bibfield  {journal} {\bibinfo  {journal} {arXiv preprint arXiv:2212.00782}\ } (\bibinfo {year} {2022})}\BibitemShut {NoStop}%
\bibitem [{\citenamefont {Luo}\ and\ \citenamefont {Clark}(2019)}]{Luo_2019}%
  \BibitemOpen
  \bibfield  {author} {\bibinfo {author} {\bibfnamefont {D.}~\bibnamefont {Luo}}\ and\ \bibinfo {author} {\bibfnamefont {B.~K.}\ \bibnamefont {Clark}},\ }\bibfield  {title} {\bibinfo {title} {Backflow transformations via neural networks for quantum many-body wave functions},\ }\bibfield  {journal} {\bibinfo  {journal} {Physical Review Letters}\ }\textbf {\bibinfo {volume} {122}},\ \href {https://doi.org/10.1103/physrevlett.122.226401} {10.1103/physrevlett.122.226401} (\bibinfo {year} {2019})\BibitemShut {NoStop}%
\bibitem [{\citenamefont {Luo}\ \emph {et~al.}(2021{\natexlab{a}})\citenamefont {Luo}, \citenamefont {Carleo}, \citenamefont {Clark},\ and\ \citenamefont {Stokes}}]{PhysRevLett.127.276402}%
  \BibitemOpen
  \bibfield  {author} {\bibinfo {author} {\bibfnamefont {D.}~\bibnamefont {Luo}}, \bibinfo {author} {\bibfnamefont {G.}~\bibnamefont {Carleo}}, \bibinfo {author} {\bibfnamefont {B.~K.}\ \bibnamefont {Clark}},\ and\ \bibinfo {author} {\bibfnamefont {J.}~\bibnamefont {Stokes}},\ }\bibfield  {title} {\bibinfo {title} {Gauge equivariant neural networks for quantum lattice gauge theories},\ }\href {https://doi.org/10.1103/PhysRevLett.127.276402} {\bibfield  {journal} {\bibinfo  {journal} {Phys. Rev. Lett.}\ }\textbf {\bibinfo {volume} {127}},\ \bibinfo {pages} {276402} (\bibinfo {year} {2021}{\natexlab{a}})}\BibitemShut {NoStop}%
\bibitem [{\citenamefont {Luo}\ \emph {et~al.}(2021{\natexlab{b}})\citenamefont {Luo}, \citenamefont {Chen}, \citenamefont {Hu}, \citenamefont {Zhao}, \citenamefont {Hur},\ and\ \citenamefont {Clark}}]{https://doi.org/10.48550/arxiv.2101.07243}%
  \BibitemOpen
  \bibfield  {author} {\bibinfo {author} {\bibfnamefont {D.}~\bibnamefont {Luo}}, \bibinfo {author} {\bibfnamefont {Z.}~\bibnamefont {Chen}}, \bibinfo {author} {\bibfnamefont {K.}~\bibnamefont {Hu}}, \bibinfo {author} {\bibfnamefont {Z.}~\bibnamefont {Zhao}}, \bibinfo {author} {\bibfnamefont {V.~M.}\ \bibnamefont {Hur}},\ and\ \bibinfo {author} {\bibfnamefont {B.~K.}\ \bibnamefont {Clark}},\ }\href {https://doi.org/10.48550/ARXIV.2101.07243} {\bibinfo {title} {Gauge invariant autoregressive neural networks for quantum lattice models}} (\bibinfo {year} {2021}{\natexlab{b}})\BibitemShut {NoStop}%
\bibitem [{\citenamefont {Luo}\ \emph {et~al.}(2022)\citenamefont {Luo}, \citenamefont {Yuan}, \citenamefont {Stokes},\ and\ \citenamefont {Clark}}]{luo2022gauge}%
  \BibitemOpen
  \bibfield  {author} {\bibinfo {author} {\bibfnamefont {D.}~\bibnamefont {Luo}}, \bibinfo {author} {\bibfnamefont {S.}~\bibnamefont {Yuan}}, \bibinfo {author} {\bibfnamefont {J.}~\bibnamefont {Stokes}},\ and\ \bibinfo {author} {\bibfnamefont {B.~K.}\ \bibnamefont {Clark}},\ }\bibfield  {title} {\bibinfo {title} {Gauge equivariant neural networks for 2+ 1d u (1) gauge theory simulations in hamiltonian formulation},\ }\href@noop {} {\bibfield  {journal} {\bibinfo  {journal} {arXiv preprint arXiv:2211.03198}\ } (\bibinfo {year} {2022})}\BibitemShut {NoStop}%
\bibitem [{\citenamefont {Glasser}\ \emph {et~al.}(2018{\natexlab{b}})\citenamefont {Glasser}, \citenamefont {Pancotti}, \citenamefont {August}, \citenamefont {Rodriguez},\ and\ \citenamefont {Cirac}}]{glasser2018neural}%
  \BibitemOpen
  \bibfield  {author} {\bibinfo {author} {\bibfnamefont {I.}~\bibnamefont {Glasser}}, \bibinfo {author} {\bibfnamefont {N.}~\bibnamefont {Pancotti}}, \bibinfo {author} {\bibfnamefont {M.}~\bibnamefont {August}}, \bibinfo {author} {\bibfnamefont {I.~D.}\ \bibnamefont {Rodriguez}},\ and\ \bibinfo {author} {\bibfnamefont {J.~I.}\ \bibnamefont {Cirac}},\ }\bibfield  {title} {\bibinfo {title} {Neural-network quantum states, string-bond states, and chiral topological states},\ }\href {https://doi.org/10.1103/PhysRevX.8.011006} {\bibfield  {journal} {\bibinfo  {journal} {Physical Review X}\ }\textbf {\bibinfo {volume} {8}},\ \bibinfo {pages} {011006} (\bibinfo {year} {2018}{\natexlab{b}})}\BibitemShut {NoStop}%
\bibitem [{\citenamefont {Pescia}\ \emph {et~al.}(2022)\citenamefont {Pescia}, \citenamefont {Han}, \citenamefont {Lovato}, \citenamefont {Lu},\ and\ \citenamefont {Carleo}}]{pescia2022neural}%
  \BibitemOpen
  \bibfield  {author} {\bibinfo {author} {\bibfnamefont {G.}~\bibnamefont {Pescia}}, \bibinfo {author} {\bibfnamefont {J.}~\bibnamefont {Han}}, \bibinfo {author} {\bibfnamefont {A.}~\bibnamefont {Lovato}}, \bibinfo {author} {\bibfnamefont {J.}~\bibnamefont {Lu}},\ and\ \bibinfo {author} {\bibfnamefont {G.}~\bibnamefont {Carleo}},\ }\bibfield  {title} {\bibinfo {title} {Neural-network quantum states for periodic systems in continuous space},\ }\href@noop {} {\bibfield  {journal} {\bibinfo  {journal} {Physical Review Research}\ }\textbf {\bibinfo {volume} {4}},\ \bibinfo {pages} {023138} (\bibinfo {year} {2022})}\BibitemShut {NoStop}%
\bibitem [{\citenamefont {Cassella}\ \emph {et~al.}(2023)\citenamefont {Cassella}, \citenamefont {Sutterud}, \citenamefont {Azadi}, \citenamefont {Drummond}, \citenamefont {Pfau}, \citenamefont {Spencer},\ and\ \citenamefont {Foulkes}}]{cassella2023discovering}%
  \BibitemOpen
  \bibfield  {author} {\bibinfo {author} {\bibfnamefont {G.}~\bibnamefont {Cassella}}, \bibinfo {author} {\bibfnamefont {H.}~\bibnamefont {Sutterud}}, \bibinfo {author} {\bibfnamefont {S.}~\bibnamefont {Azadi}}, \bibinfo {author} {\bibfnamefont {N.}~\bibnamefont {Drummond}}, \bibinfo {author} {\bibfnamefont {D.}~\bibnamefont {Pfau}}, \bibinfo {author} {\bibfnamefont {J.~S.}\ \bibnamefont {Spencer}},\ and\ \bibinfo {author} {\bibfnamefont {W.~M.~C.}\ \bibnamefont {Foulkes}},\ }\bibfield  {title} {\bibinfo {title} {Discovering quantum phase transitions with fermionic neural networks},\ }\href@noop {} {\bibfield  {journal} {\bibinfo  {journal} {Physical Review Letters}\ }\textbf {\bibinfo {volume} {130}},\ \bibinfo {pages} {036401} (\bibinfo {year} {2023})}\BibitemShut {NoStop}%
\bibitem [{\citenamefont {Pescia}\ \emph {et~al.}(2023)\citenamefont {Pescia}, \citenamefont {Nys}, \citenamefont {Kim}, \citenamefont {Lovato},\ and\ \citenamefont {Carleo}}]{pescia2023message}%
  \BibitemOpen
  \bibfield  {author} {\bibinfo {author} {\bibfnamefont {G.}~\bibnamefont {Pescia}}, \bibinfo {author} {\bibfnamefont {J.}~\bibnamefont {Nys}}, \bibinfo {author} {\bibfnamefont {J.}~\bibnamefont {Kim}}, \bibinfo {author} {\bibfnamefont {A.}~\bibnamefont {Lovato}},\ and\ \bibinfo {author} {\bibfnamefont {G.}~\bibnamefont {Carleo}},\ }\bibfield  {title} {\bibinfo {title} {Message-passing neural quantum states for the homogeneous electron gas},\ }\href@noop {} {\bibfield  {journal} {\bibinfo  {journal} {arXiv preprint arXiv:2305.07240}\ } (\bibinfo {year} {2023})}\BibitemShut {NoStop}%
\bibitem [{\citenamefont {Kim}\ \emph {et~al.}(2023)\citenamefont {Kim}, \citenamefont {Pescia}, \citenamefont {Fore}, \citenamefont {Nys}, \citenamefont {Carleo}, \citenamefont {Gandolfi}, \citenamefont {Hjorth-Jensen},\ and\ \citenamefont {Lovato}}]{kim2023neural}%
  \BibitemOpen
  \bibfield  {author} {\bibinfo {author} {\bibfnamefont {J.}~\bibnamefont {Kim}}, \bibinfo {author} {\bibfnamefont {G.}~\bibnamefont {Pescia}}, \bibinfo {author} {\bibfnamefont {B.}~\bibnamefont {Fore}}, \bibinfo {author} {\bibfnamefont {J.}~\bibnamefont {Nys}}, \bibinfo {author} {\bibfnamefont {G.}~\bibnamefont {Carleo}}, \bibinfo {author} {\bibfnamefont {S.}~\bibnamefont {Gandolfi}}, \bibinfo {author} {\bibfnamefont {M.}~\bibnamefont {Hjorth-Jensen}},\ and\ \bibinfo {author} {\bibfnamefont {A.}~\bibnamefont {Lovato}},\ }\bibfield  {title} {\bibinfo {title} {Neural-network quantum states for ultra-cold fermi gases},\ }\href@noop {} {\bibfield  {journal} {\bibinfo  {journal} {arXiv preprint arXiv:2305.08831}\ } (\bibinfo {year} {2023})}\BibitemShut {NoStop}%
\bibitem [{\citenamefont {Lou}\ \emph {et~al.}(2023)\citenamefont {Lou}, \citenamefont {Sutterud}, \citenamefont {Cassella}, \citenamefont {Foulkes}, \citenamefont {Knolle}, \citenamefont {Pfau},\ and\ \citenamefont {Spencer}}]{lou2023neural}%
  \BibitemOpen
  \bibfield  {author} {\bibinfo {author} {\bibfnamefont {W.~T.}\ \bibnamefont {Lou}}, \bibinfo {author} {\bibfnamefont {H.}~\bibnamefont {Sutterud}}, \bibinfo {author} {\bibfnamefont {G.}~\bibnamefont {Cassella}}, \bibinfo {author} {\bibfnamefont {W.}~\bibnamefont {Foulkes}}, \bibinfo {author} {\bibfnamefont {J.}~\bibnamefont {Knolle}}, \bibinfo {author} {\bibfnamefont {D.}~\bibnamefont {Pfau}},\ and\ \bibinfo {author} {\bibfnamefont {J.~S.}\ \bibnamefont {Spencer}},\ }\bibfield  {title} {\bibinfo {title} {Neural wave functions for superfluids},\ }\href@noop {} {\bibfield  {journal} {\bibinfo  {journal} {arXiv preprint arXiv:2305.06989}\ } (\bibinfo {year} {2023})}\BibitemShut {NoStop}%
\bibitem [{\citenamefont {Entwistle}\ \emph {et~al.}(2023)\citenamefont {Entwistle}, \citenamefont {Sch{\"a}tzle}, \citenamefont {Erdman}, \citenamefont {Hermann},\ and\ \citenamefont {No{\'e}}}]{entwistle2023electronic}%
  \BibitemOpen
  \bibfield  {author} {\bibinfo {author} {\bibfnamefont {M.~T.}\ \bibnamefont {Entwistle}}, \bibinfo {author} {\bibfnamefont {Z.}~\bibnamefont {Sch{\"a}tzle}}, \bibinfo {author} {\bibfnamefont {P.~A.}\ \bibnamefont {Erdman}}, \bibinfo {author} {\bibfnamefont {J.}~\bibnamefont {Hermann}},\ and\ \bibinfo {author} {\bibfnamefont {F.}~\bibnamefont {No{\'e}}},\ }\bibfield  {title} {\bibinfo {title} {Electronic excited states in deep variational monte carlo},\ }\href@noop {} {\bibfield  {journal} {\bibinfo  {journal} {Nature Communications}\ }\textbf {\bibinfo {volume} {14}},\ \bibinfo {pages} {274} (\bibinfo {year} {2023})}\BibitemShut {NoStop}%
\bibitem [{\citenamefont {Wilson}\ \emph {et~al.}(2022)\citenamefont {Wilson}, \citenamefont {Moroni}, \citenamefont {Holzmann}, \citenamefont {Gao}, \citenamefont {Wudarski}, \citenamefont {Vegge},\ and\ \citenamefont {Bhowmik}}]{wilson2022wave}%
  \BibitemOpen
  \bibfield  {author} {\bibinfo {author} {\bibfnamefont {M.}~\bibnamefont {Wilson}}, \bibinfo {author} {\bibfnamefont {S.}~\bibnamefont {Moroni}}, \bibinfo {author} {\bibfnamefont {M.}~\bibnamefont {Holzmann}}, \bibinfo {author} {\bibfnamefont {N.}~\bibnamefont {Gao}}, \bibinfo {author} {\bibfnamefont {F.}~\bibnamefont {Wudarski}}, \bibinfo {author} {\bibfnamefont {T.}~\bibnamefont {Vegge}},\ and\ \bibinfo {author} {\bibfnamefont {A.}~\bibnamefont {Bhowmik}},\ }\bibfield  {title} {\bibinfo {title} {Wave function ansatz (but periodic) networks and the homogeneous electron gas},\ }\href@noop {} {\bibfield  {journal} {\bibinfo  {journal} {arXiv preprint arXiv:2202.04622}\ } (\bibinfo {year} {2022})}\BibitemShut {NoStop}%
\bibitem [{\citenamefont {Li}\ \emph {et~al.}(2022)\citenamefont {Li}, \citenamefont {Li},\ and\ \citenamefont {Chen}}]{li2022ab}%
  \BibitemOpen
  \bibfield  {author} {\bibinfo {author} {\bibfnamefont {X.}~\bibnamefont {Li}}, \bibinfo {author} {\bibfnamefont {Z.}~\bibnamefont {Li}},\ and\ \bibinfo {author} {\bibfnamefont {J.}~\bibnamefont {Chen}},\ }\bibfield  {title} {\bibinfo {title} {Ab initio calculation of real solids via neural network ansatz},\ }\href@noop {} {\bibfield  {journal} {\bibinfo  {journal} {Nature Communications}\ }\textbf {\bibinfo {volume} {13}},\ \bibinfo {pages} {7895} (\bibinfo {year} {2022})}\BibitemShut {NoStop}%
\bibitem [{\citenamefont {Scherbela}\ \emph {et~al.}(2022)\citenamefont {Scherbela}, \citenamefont {Reisenhofer}, \citenamefont {Gerard}, \citenamefont {Marquetand},\ and\ \citenamefont {Grohs}}]{scherbela2022solving}%
  \BibitemOpen
  \bibfield  {author} {\bibinfo {author} {\bibfnamefont {M.}~\bibnamefont {Scherbela}}, \bibinfo {author} {\bibfnamefont {R.}~\bibnamefont {Reisenhofer}}, \bibinfo {author} {\bibfnamefont {L.}~\bibnamefont {Gerard}}, \bibinfo {author} {\bibfnamefont {P.}~\bibnamefont {Marquetand}},\ and\ \bibinfo {author} {\bibfnamefont {P.}~\bibnamefont {Grohs}},\ }\bibfield  {title} {\bibinfo {title} {Solving the electronic schr{\"o}dinger equation for multiple nuclear geometries with weight-sharing deep neural networks},\ }\href@noop {} {\bibfield  {journal} {\bibinfo  {journal} {Nature Computational Science}\ }\textbf {\bibinfo {volume} {2}},\ \bibinfo {pages} {331} (\bibinfo {year} {2022})}\BibitemShut {NoStop}%
\bibitem [{\citenamefont {Adams}\ \emph {et~al.}(2021)\citenamefont {Adams}, \citenamefont {Carleo}, \citenamefont {Lovato},\ and\ \citenamefont {Rocco}}]{adams2021variational}%
  \BibitemOpen
  \bibfield  {author} {\bibinfo {author} {\bibfnamefont {C.}~\bibnamefont {Adams}}, \bibinfo {author} {\bibfnamefont {G.}~\bibnamefont {Carleo}}, \bibinfo {author} {\bibfnamefont {A.}~\bibnamefont {Lovato}},\ and\ \bibinfo {author} {\bibfnamefont {N.}~\bibnamefont {Rocco}},\ }\bibfield  {title} {\bibinfo {title} {Variational monte carlo calculations of $a\ensuremath{\le}4$ nuclei with an artificial neural-network correlator ansatz},\ }\href@noop {} {\bibfield  {journal} {\bibinfo  {journal} {Physical Review Letters}\ }\textbf {\bibinfo {volume} {127}},\ \bibinfo {pages} {022502} (\bibinfo {year} {2021})}\BibitemShut {NoStop}%
\bibitem [{\citenamefont {Smith}\ \emph {et~al.}(2024)\citenamefont {Smith}, \citenamefont {Chen}, \citenamefont {Levy}, \citenamefont {Yang}, \citenamefont {Morales},\ and\ \citenamefont {Zhang}}]{smith2024ground}%
  \BibitemOpen
  \bibfield  {author} {\bibinfo {author} {\bibfnamefont {C.}~\bibnamefont {Smith}}, \bibinfo {author} {\bibfnamefont {Y.}~\bibnamefont {Chen}}, \bibinfo {author} {\bibfnamefont {R.}~\bibnamefont {Levy}}, \bibinfo {author} {\bibfnamefont {Y.}~\bibnamefont {Yang}}, \bibinfo {author} {\bibfnamefont {M.~A.}\ \bibnamefont {Morales}},\ and\ \bibinfo {author} {\bibfnamefont {S.}~\bibnamefont {Zhang}},\ }\bibfield  {title} {\bibinfo {title} {Ground state phases of the two-dimension electron gas with a unified variational approach},\ }\href@noop {} {\bibfield  {journal} {\bibinfo  {journal} {arXiv preprint arXiv:2405.19397}\ } (\bibinfo {year} {2024})}\BibitemShut {NoStop}%
\bibitem [{\citenamefont {Luo}\ \emph {et~al.}(2023)\citenamefont {Luo}, \citenamefont {Dai},\ and\ \citenamefont {Fu}}]{luo2023pairing}%
  \BibitemOpen
  \bibfield  {author} {\bibinfo {author} {\bibfnamefont {D.}~\bibnamefont {Luo}}, \bibinfo {author} {\bibfnamefont {D.~D.}\ \bibnamefont {Dai}},\ and\ \bibinfo {author} {\bibfnamefont {L.}~\bibnamefont {Fu}},\ }\bibfield  {title} {\bibinfo {title} {Pairing-based graph neural network for simulating quantum materials},\ }\href@noop {} {\bibfield  {journal} {\bibinfo  {journal} {arXiv preprint arXiv:2311.02143}\ } (\bibinfo {year} {2023})}\BibitemShut {NoStop}%
\bibitem [{\citenamefont {Teng}\ \emph {et~al.}(2024)\citenamefont {Teng}, \citenamefont {Dai},\ and\ \citenamefont {Fu}}]{teng2024solving}%
  \BibitemOpen
  \bibfield  {author} {\bibinfo {author} {\bibfnamefont {Y.}~\bibnamefont {Teng}}, \bibinfo {author} {\bibfnamefont {D.~D.}\ \bibnamefont {Dai}},\ and\ \bibinfo {author} {\bibfnamefont {L.}~\bibnamefont {Fu}},\ }\href {https://arxiv.org/abs/2412.00618} {\bibinfo {title} {Solving and visualizing fractional quantum hall wavefunctions with neural network}} (\bibinfo {year} {2024}),\ \Eprint {https://arxiv.org/abs/2412.00618} {arXiv:2412.00618 [cond-mat.str-el]} \BibitemShut {NoStop}%
\bibitem [{\citenamefont {Qian}\ \emph {et~al.}(2024)\citenamefont {Qian}, \citenamefont {Zhao}, \citenamefont {Zhang}, \citenamefont {Xiang}, \citenamefont {Li},\ and\ \citenamefont {Chen}}]{qian2024taming}%
  \BibitemOpen
  \bibfield  {author} {\bibinfo {author} {\bibfnamefont {Y.}~\bibnamefont {Qian}}, \bibinfo {author} {\bibfnamefont {T.}~\bibnamefont {Zhao}}, \bibinfo {author} {\bibfnamefont {J.}~\bibnamefont {Zhang}}, \bibinfo {author} {\bibfnamefont {T.}~\bibnamefont {Xiang}}, \bibinfo {author} {\bibfnamefont {X.}~\bibnamefont {Li}},\ and\ \bibinfo {author} {\bibfnamefont {J.}~\bibnamefont {Chen}},\ }\href {https://arxiv.org/abs/2412.14795} {\bibinfo {title} {Taming landau level mixing in fractional quantum hall states with deep learning}} (\bibinfo {year} {2024}),\ \Eprint {https://arxiv.org/abs/2412.14795} {arXiv:2412.14795 [cond-mat.str-el]} \BibitemShut {NoStop}%
\bibitem [{\citenamefont {Li}\ \emph {et~al.}(2024)\citenamefont {Li}, \citenamefont {Qian}, \citenamefont {Ren}, \citenamefont {Xu},\ and\ \citenamefont {Chen}}]{li2024emergent}%
  \BibitemOpen
  \bibfield  {author} {\bibinfo {author} {\bibfnamefont {X.}~\bibnamefont {Li}}, \bibinfo {author} {\bibfnamefont {Y.}~\bibnamefont {Qian}}, \bibinfo {author} {\bibfnamefont {W.}~\bibnamefont {Ren}}, \bibinfo {author} {\bibfnamefont {Y.}~\bibnamefont {Xu}},\ and\ \bibinfo {author} {\bibfnamefont {J.}~\bibnamefont {Chen}},\ }\bibfield  {title} {\bibinfo {title} {Emergent wigner phases in moir\'e superlattice from deep learning},\ }\href@noop {} {\bibfield  {journal} {\bibinfo  {journal} {arXiv preprint arXiv:2406.11134}\ } (\bibinfo {year} {2024})}\BibitemShut {NoStop}%
\bibitem [{\citenamefont {Luo}\ \emph {et~al.}(2024)\citenamefont {Luo}, \citenamefont {Dai},\ and\ \citenamefont {Fu}}]{Luo2024NNMoire}%
  \BibitemOpen
  \bibfield  {author} {\bibinfo {author} {\bibfnamefont {D.}~\bibnamefont {Luo}}, \bibinfo {author} {\bibfnamefont {D.~D.}\ \bibnamefont {Dai}},\ and\ \bibinfo {author} {\bibfnamefont {L.}~\bibnamefont {Fu}},\ }\bibfield  {title} {\bibinfo {title} {Simulating moiré quantum matter with neural network},\ }\href@noop {} {\bibfield  {journal} {\bibinfo  {journal} {arXiv preprint arXiv:2406.17645}\ } (\bibinfo {year} {2024})}\BibitemShut {NoStop}%
\bibitem [{\citenamefont {Geier}\ \emph {et~al.}(2025)\citenamefont {Geier}, \citenamefont {Nazaryan}, \citenamefont {Zaklama},\ and\ \citenamefont {Fu}}]{geier2025attention}%
  \BibitemOpen
  \bibfield  {author} {\bibinfo {author} {\bibfnamefont {M.}~\bibnamefont {Geier}}, \bibinfo {author} {\bibfnamefont {K.}~\bibnamefont {Nazaryan}}, \bibinfo {author} {\bibfnamefont {T.}~\bibnamefont {Zaklama}},\ and\ \bibinfo {author} {\bibfnamefont {L.}~\bibnamefont {Fu}},\ }\href {https://arxiv.org/abs/2502.05383} {\bibinfo {title} {Is attention all you need to solve the correlated electron problem?}} (\bibinfo {year} {2025}),\ \Eprint {https://arxiv.org/abs/2502.05383} {arXiv:2502.05383 [cond-mat.str-el]} \BibitemShut {NoStop}%
\bibitem [{\citenamefont {Wu}\ \emph {et~al.}(2019)\citenamefont {Wu}, \citenamefont {Lovorn}, \citenamefont {Tutuc}, \citenamefont {Martin},\ and\ \citenamefont {MacDonald}}]{wu2019topological}%
  \BibitemOpen
  \bibfield  {author} {\bibinfo {author} {\bibfnamefont {F.}~\bibnamefont {Wu}}, \bibinfo {author} {\bibfnamefont {T.}~\bibnamefont {Lovorn}}, \bibinfo {author} {\bibfnamefont {E.}~\bibnamefont {Tutuc}}, \bibinfo {author} {\bibfnamefont {I.}~\bibnamefont {Martin}},\ and\ \bibinfo {author} {\bibfnamefont {A.}~\bibnamefont {MacDonald}},\ }\bibfield  {title} {\bibinfo {title} {Topological insulators in twisted transition metal dichalcogenide homobilayers},\ }\href {https://doi.org/10.1103/PhysRevLett.122.086402} {\bibfield  {journal} {\bibinfo  {journal} {Physical review letters}\ }\textbf {\bibinfo {volume} {122}},\ \bibinfo {pages} {086402} (\bibinfo {year} {2019})}\BibitemShut {NoStop}%
\bibitem [{\citenamefont {Wu}\ \emph {et~al.}(2013)\citenamefont {Wu}, \citenamefont {Regnault},\ and\ \citenamefont {Bernevig}}]{wu2013bloch}%
  \BibitemOpen
  \bibfield  {author} {\bibinfo {author} {\bibfnamefont {Y.-L.}\ \bibnamefont {Wu}}, \bibinfo {author} {\bibfnamefont {N.}~\bibnamefont {Regnault}},\ and\ \bibinfo {author} {\bibfnamefont {B.~A.}\ \bibnamefont {Bernevig}},\ }\bibfield  {title} {\bibinfo {title} {Bloch model wave functions and pseudopotentials for all fractional chern insulators},\ }\href@noop {} {\bibfield  {journal} {\bibinfo  {journal} {Physical review letters}\ }\textbf {\bibinfo {volume} {110}},\ \bibinfo {pages} {106802} (\bibinfo {year} {2013})}\BibitemShut {NoStop}%
\bibitem [{\citenamefont {Sorella}(1998)}]{sorella1998green}%
  \BibitemOpen
  \bibfield  {author} {\bibinfo {author} {\bibfnamefont {S.}~\bibnamefont {Sorella}},\ }\bibfield  {title} {\bibinfo {title} {Green function monte carlo with stochastic reconfiguration},\ }\href@noop {} {\bibfield  {journal} {\bibinfo  {journal} {Physical review letters}\ }\textbf {\bibinfo {volume} {80}},\ \bibinfo {pages} {4558} (\bibinfo {year} {1998})}\BibitemShut {NoStop}%
\bibitem [{\citenamefont {Onishi}\ and\ \citenamefont {Fu}(2024{\natexlab{a}})}]{onishi2024quantum}%
  \BibitemOpen
  \bibfield  {author} {\bibinfo {author} {\bibfnamefont {Y.}~\bibnamefont {Onishi}}\ and\ \bibinfo {author} {\bibfnamefont {L.}~\bibnamefont {Fu}},\ }\bibfield  {title} {\bibinfo {title} {Quantum weight},\ }\href@noop {} {\bibfield  {journal} {\bibinfo  {journal} {arXiv preprint arXiv:2401.13847}\ } (\bibinfo {year} {2024}{\natexlab{a}})}\BibitemShut {NoStop}%
\bibitem [{\citenamefont {Onishi}\ and\ \citenamefont {Fu}(2024{\natexlab{b}})}]{onishi2023quantum}%
  \BibitemOpen
  \bibfield  {author} {\bibinfo {author} {\bibfnamefont {Y.}~\bibnamefont {Onishi}}\ and\ \bibinfo {author} {\bibfnamefont {L.}~\bibnamefont {Fu}},\ }\bibfield  {title} {\bibinfo {title} {{Fundamental Bound on Topological Gap}},\ }\href {https://doi.org/10.1103/PhysRevX.14.011052} {\bibfield  {journal} {\bibinfo  {journal} {Phys. Rev. X}\ }\textbf {\bibinfo {volume} {14}},\ \bibinfo {pages} {011052} (\bibinfo {year} {2024}{\natexlab{b}})}\BibitemShut {NoStop}%
\bibitem [{\citenamefont {Zaklama}\ \emph {et~al.}(2024)\citenamefont {Zaklama}, \citenamefont {Luo},\ and\ \citenamefont {Fu}}]{zaklama2024structure}%
  \BibitemOpen
  \bibfield  {author} {\bibinfo {author} {\bibfnamefont {T.}~\bibnamefont {Zaklama}}, \bibinfo {author} {\bibfnamefont {D.}~\bibnamefont {Luo}},\ and\ \bibinfo {author} {\bibfnamefont {L.}~\bibnamefont {Fu}},\ }\bibfield  {title} {\bibinfo {title} {Structure factor and topological bound of twisted bilayer semiconductors at fractional fillings},\ }\href@noop {} {\bibfield  {journal} {\bibinfo  {journal} {arXiv preprint arXiv:2411.03496}\ } (\bibinfo {year} {2024})}\BibitemShut {NoStop}%
\bibitem [{\citenamefont {Onishi}\ and\ \citenamefont {Fu}(2024{\natexlab{c}})}]{onishi2024StructureFact}%
  \BibitemOpen
  \bibfield  {author} {\bibinfo {author} {\bibfnamefont {Y.}~\bibnamefont {Onishi}}\ and\ \bibinfo {author} {\bibfnamefont {L.}~\bibnamefont {Fu}},\ }\bibfield  {title} {\bibinfo {title} {Topological bound on structure factor},\ }\href@noop {} {\bibfield  {journal} {\bibinfo  {journal} {arXiv preprint arXiv:2406.18654}\ } (\bibinfo {year} {2024}{\natexlab{c}})}\BibitemShut {NoStop}%
\bibitem [{\citenamefont {Kang}\ \emph {et~al.}(2024)\citenamefont {Kang}, \citenamefont {Shen}, \citenamefont {Qiu}, \citenamefont {Watanabe}, \citenamefont {Taniguchi}, \citenamefont {Shan},\ and\ \citenamefont {Mak}}]{kang2024observationMoTe2}%
  \BibitemOpen
  \bibfield  {author} {\bibinfo {author} {\bibfnamefont {K.}~\bibnamefont {Kang}}, \bibinfo {author} {\bibfnamefont {B.}~\bibnamefont {Shen}}, \bibinfo {author} {\bibfnamefont {Y.}~\bibnamefont {Qiu}}, \bibinfo {author} {\bibfnamefont {K.}~\bibnamefont {Watanabe}}, \bibinfo {author} {\bibfnamefont {T.}~\bibnamefont {Taniguchi}}, \bibinfo {author} {\bibfnamefont {J.}~\bibnamefont {Shan}},\ and\ \bibinfo {author} {\bibfnamefont {K.~F.}\ \bibnamefont {Mak}},\ }\bibfield  {title} {\bibinfo {title} {Observation of the fractional quantum spin hall effect in moir$\backslash$'e mote2},\ }\href@noop {} {\bibfield  {journal} {\bibinfo  {journal} {arXiv preprint arXiv:2402.03294}\ } (\bibinfo {year} {2024})}\BibitemShut {NoStop}%
\bibitem [{\citenamefont {Yu}\ \emph {et~al.}(2024{\natexlab{b}})\citenamefont {Yu}, \citenamefont {Herzog-Arbeitman}, \citenamefont {Wang}, \citenamefont {Vafek}, \citenamefont {Bernevig},\ and\ \citenamefont {Regnault}}]{Yu2023fractional}%
  \BibitemOpen
  \bibfield  {author} {\bibinfo {author} {\bibfnamefont {J.}~\bibnamefont {Yu}}, \bibinfo {author} {\bibfnamefont {J.}~\bibnamefont {Herzog-Arbeitman}}, \bibinfo {author} {\bibfnamefont {M.}~\bibnamefont {Wang}}, \bibinfo {author} {\bibfnamefont {O.}~\bibnamefont {Vafek}}, \bibinfo {author} {\bibfnamefont {B.~A.}\ \bibnamefont {Bernevig}},\ and\ \bibinfo {author} {\bibfnamefont {N.}~\bibnamefont {Regnault}},\ }\bibfield  {title} {\bibinfo {title} {Fractional chern insulators versus nonmagnetic states in twisted bilayer ${\mathrm{mote}}_{2}$},\ }\href {https://doi.org/10.1103/PhysRevB.109.045147} {\bibfield  {journal} {\bibinfo  {journal} {Phys. Rev. B}\ }\textbf {\bibinfo {volume} {109}},\ \bibinfo {pages} {045147} (\bibinfo {year} {2024}{\natexlab{b}})}\BibitemShut {NoStop}%
\bibitem [{\citenamefont {Wilhelm}\ \emph {et~al.}(2021)\citenamefont {Wilhelm}, \citenamefont {Lang},\ and\ \citenamefont {L{\"a}uchli}}]{wilhelm2021interplay}%
  \BibitemOpen
  \bibfield  {author} {\bibinfo {author} {\bibfnamefont {P.}~\bibnamefont {Wilhelm}}, \bibinfo {author} {\bibfnamefont {T.~C.}\ \bibnamefont {Lang}},\ and\ \bibinfo {author} {\bibfnamefont {A.~M.}\ \bibnamefont {L{\"a}uchli}},\ }\bibfield  {title} {\bibinfo {title} {Interplay of fractional chern insulator and charge density wave phases in twisted bilayer graphene},\ }\href@noop {} {\bibfield  {journal} {\bibinfo  {journal} {Physical Review B}\ }\textbf {\bibinfo {volume} {103}},\ \bibinfo {pages} {125406} (\bibinfo {year} {2021})}\BibitemShut {NoStop}%
\bibitem [{\citenamefont {Klambauer}\ \emph {et~al.}(2017)\citenamefont {Klambauer}, \citenamefont {Unterthiner}, \citenamefont {Mayr},\ and\ \citenamefont {Hochreiter}}]{klambauer2017self}%
  \BibitemOpen
  \bibfield  {author} {\bibinfo {author} {\bibfnamefont {G.}~\bibnamefont {Klambauer}}, \bibinfo {author} {\bibfnamefont {T.}~\bibnamefont {Unterthiner}}, \bibinfo {author} {\bibfnamefont {A.}~\bibnamefont {Mayr}},\ and\ \bibinfo {author} {\bibfnamefont {S.}~\bibnamefont {Hochreiter}},\ }\bibfield  {title} {\bibinfo {title} {Self-normalizing neural networks},\ }\href@noop {} {\bibfield  {journal} {\bibinfo  {journal} {Advances in neural information processing systems}\ }\textbf {\bibinfo {volume} {30}} (\bibinfo {year} {2017})}\BibitemShut {NoStop}%
\bibitem [{\citenamefont {Bradbury}\ \emph {et~al.}(2018)\citenamefont {Bradbury}, \citenamefont {Frostig}, \citenamefont {Hawkins}, \citenamefont {Johnson}, \citenamefont {Leary}, \citenamefont {Maclaurin}, \citenamefont {Necula}, \citenamefont {Paszke}, \citenamefont {Vander{P}las}, \citenamefont {Wanderman-{M}ilne},\ and\ \citenamefont {Zhang}}]{jax2018github}%
  \BibitemOpen
  \bibfield  {author} {\bibinfo {author} {\bibfnamefont {J.}~\bibnamefont {Bradbury}}, \bibinfo {author} {\bibfnamefont {R.}~\bibnamefont {Frostig}}, \bibinfo {author} {\bibfnamefont {P.}~\bibnamefont {Hawkins}}, \bibinfo {author} {\bibfnamefont {M.~J.}\ \bibnamefont {Johnson}}, \bibinfo {author} {\bibfnamefont {C.}~\bibnamefont {Leary}}, \bibinfo {author} {\bibfnamefont {D.}~\bibnamefont {Maclaurin}}, \bibinfo {author} {\bibfnamefont {G.}~\bibnamefont {Necula}}, \bibinfo {author} {\bibfnamefont {A.}~\bibnamefont {Paszke}}, \bibinfo {author} {\bibfnamefont {J.}~\bibnamefont {Vander{P}las}}, \bibinfo {author} {\bibfnamefont {S.}~\bibnamefont {Wanderman-{M}ilne}},\ and\ \bibinfo {author} {\bibfnamefont {Q.}~\bibnamefont {Zhang}},\ }\href {http://github.com/google/jax} {\bibinfo {title} {{JAX}: composable transformations of {P}ython+{N}um{P}y programs}} (\bibinfo {year} {2018})\BibitemShut {NoStop}%
\end{thebibliography}%

\appendix

\clearpage

\onecolumngrid
\begin{center}
	\noindent\textbf{Supplementary Materials for Solving Fractional Electron States in Twisted Bilayer MoTe$_2$ with Deep Neural Network}
	\bigskip
		
	\noindent\textbf{\large{}}
\end{center}

\section{I. Details on Exact Diagonalization}

In order to solve the Many-Body System using the exact diagonlization method, we use the continuum description of holes in a single valley, which has been widely used to describe moiré systems \cite{wu2019topological}. In the continuum description, the edges of the valence band within each TMD layer are located at the $K$ and $K$' points of the Brillouin zone. At these points, the particles carry a large effective mass in the range of $\frac{m_e}{2}$ to $m_e$. Strong spin-orbit coupling locks the spin in opposite directions of holes at K and K', thereby reducing the spin and valley degrees of freedom into a single spin degree of freedom. In AA-stacked TMD homobilayers (TMD monolayers stacked with no offset or twist), holes at a given valley have the same spin in both layers, which creates direct spin-conservation along with intra-valley and inter-layer tunneling. The rotation of one layer with respect to another creates a moiré pattern of the combined layers, introducing an intra-layer superlattice potential and inter-layer tunneling that vary with superlattice periodicity. The dispersion relation for the low energy holes is also modified, and following the simple gauge transformation of $U(\mathbf{r})\hat{H}U^{\dagger}(\mathbf{r})$ for $U=\begin{pmatrix}
    e^{i\kappa_+ r} & 0 \\ 0 & e^{i\kappa_- r}
\end{pmatrix}$ to the layer basis, we can rewrite our interacting continuum Hamiltonian in Eq. \ref{eq:Ham} as 

\begin{equation}
    \mathcal{H}_{\uparrow} = \begin{pmatrix}
  \frac{\hbar^2(-i\nabla_{-\kappa_+)^2}}{2m}+V_1(\mathbf{r}) & t(\mathbf{r}) \\ t^\dagger(\mathbf{r}) & \frac{\hbar^2(-i\nabla_{-\kappa_-)^2}}{2m}+V_2(\mathbf{r})
\end{pmatrix},
\end{equation}

where the origin is midway along the rotation axis of the two layers, and $\mathcal{H}_{\downarrow}$ is the time reversal conjugate. The corners of the moiré Brillouin zone (MBZ) are $\mathbf{\kappa}_{\pm}=\frac{4\pi}{3a_M}\left(\frac{\sqrt{3}}{2},\pm\frac{1}{2}\right)$, with moiré period $a_M=\frac{a_0}{2\sin(\frac{\theta}{2})}$, for monolayer lattice constant $a_0$ and twist angle $\theta$. $V_{l}(\mathbf{r})$ and $t(\mathbf{r})$ represent the intralayer moiré potential and interlayer tunneling respectively, for layer index $l$. By taking into account the bilayer system's $D_3$ point group symmetry and fourier expanding both functions to their lowest harmonics, we can express these terms in the following way:

\begin{align}
    \begin{split}
        V_{l}(\mathbf{r}) & =  -2V\sum_{i=1,3,5} \cos(\mathbf{g}_i+\phi_l) \\
        t(\mathbf{r}) & =w\left(1 + e^{-i\mathbf{g}_2\cdot\mathbf{r}} + e^{-i\mathbf{g}_3\cdot\mathbf{r}}\right).
    \end{split}
\end{align}

Here, the moiré reciprocal lattice vectors are $\mathbf{g}_i=\frac{4\pi}{\sqrt{3}a_M}\left(\cos(\frac{\pi(i-1)}{3}),\sin(\frac{\pi(i-1)}{3})\right)$ for $i=1,\dots,6$, and $\phi_1=-\phi_2=\phi$. Since this Hamiltonian is defined with a charge neutral vacuum, we choose to write the continuum model Hamiltonian exclusively in terms of hole operators: 

\begin{equation}
    \mathbf{H}_0 = \sum_{\sigma=\uparrow,\downarrow} \int d\mathbf{r} \psi_{\sigma}^\dagger \mathcal{H}_\sigma \psi_\sigma,
\end{equation}

where $\psi_\sigma^\dagger$ creates a hole in the valence band, which bounds from below the $\mathbf{H}_0$ energy spectrum. 
When the moiré period $a_M$ is far greater than the monolayer lattice constant, the single-particle moiré band structure is accurately described by the continuum model, which approximates effective mass in the semiconductor band edge, and includes a slowly varying effective periodic potential that arises from the band edge modulation within the moiré unit cell.

The full continuum model Hamiltonian including electron-electron interaction is given by
\begin{equation}
\begin{split}
    \bm{H} &= \bm{H}_0 + \bm{V}, \\
    \bm{V} &= \frac{1}{2}\sum_{\sigma,\sigma'}\int d\mathbf{r}d\mathbf{r'}\psi_\sigma^\dagger(\mathbf{r})\psi_{\sigma'}^\dagger(\mathbf{r'})V(\mathbf{r}-\mathbf{r'})\psi_{\sigma'}(\mathbf{r'})\psi_{\sigma}(\mathbf{r}),
\end{split}
\end{equation}
where $V(\mathbf{r})=\frac{e^2}{\epsilon r}$ is the long range Coulomb interaction. By diagonalizing the one-body Hamiltonian $\bm{H}_0$, we obtain the band dispersion and Bloch wavefunction. 
Then, $\bm{H}$ can be rewritten in Bloch band basis as follows: 
\begin{equation}
    \bm{\Tilde{H}}=\sum_{\sigma,n,\mathbf{k}}\epsilon_{\sigma,n}(\mathbf{k})c_{\sigma,n,\mathbf{k}}^\dagger c_{\sigma,n,\mathbf{k}}
    +\frac{1}{2} \sum_{\substack{\sigma,\sigma'; \mathbf{k'_1}\\ n_1,n_2,n_3,n_4;\\ \mathbf{k'_2}\mathbf{k_1}, \mathbf{k_2}}} V_{\mathbf{k'_1}\mathbf{k'_2}\mathbf{k_1}\mathbf{k_2};\sigma\sigma'}^{n_1n_2n_3n_4} c^\dagger_{\sigma_1,n_1,\mathbf{k'_1}} c^\dagger_{\sigma_2,n_3,\mathbf{k'_2}} c_{\sigma_2,n_4,\mathbf{k_2}} c_{\sigma_1,n_2,\mathbf{k_1}} \label{Hband}
\end{equation}

where $c_{\sigma,n,\mathbf{k}}^\dagger$ creates a hole in a Bloch state at spin/valley $\sigma$, band $n$,  crystal momentum $\mathbf{k}$, which has a corresponding single-particle energy $\epsilon_{\sigma,n}(\mathbf{k})$. The Coulomb interaction matrix elements take on the Bloch basis representation $V_{\mathbf{k'_1}\mathbf{k'_2}\mathbf{k_1}\mathbf{k_2};\sigma\sigma'}^{n_1n_2n_3n_4} \equiv \langle \mathbf{k'_1}n_1\sigma;\mathbf{k'_2}n_2\sigma'|\hat{V}|\mathbf{k_1}n_3\sigma;\mathbf{k_2}n_4\sigma' \rangle$.

Our ED calculation uses the charge-$U(1)$, spin-$U(1)$, and spatial translational symmetries to diagonalize within the common eigenspace of $N_e$, $S_z$, and center of mass (CoM) crystal momentum. We neglect the spin degrees of freedom as we restrict the discussion to fully spin polarized regime at various fractional fillings. 
All clusters used possess $C_6$ point group symmetry, rendering the system isotropic as a result of its triangular $2d$ structure. This band projection neglects band mixing with higher bands, which is accurate when the ratio of the characteristic Coulomb energy $\frac{e^2}{\epsilon a_M}$ to the moiré band gap is sufficiently small. Previous studies have shown that band mixing is relevant for this system \cite{abouelkomsan2024band,Yu2023fractional}, as our neural Block wavefunction reinforces in this work. 

\begin{figure*}[t]
\centering
        \includegraphics[width=0.6\linewidth]{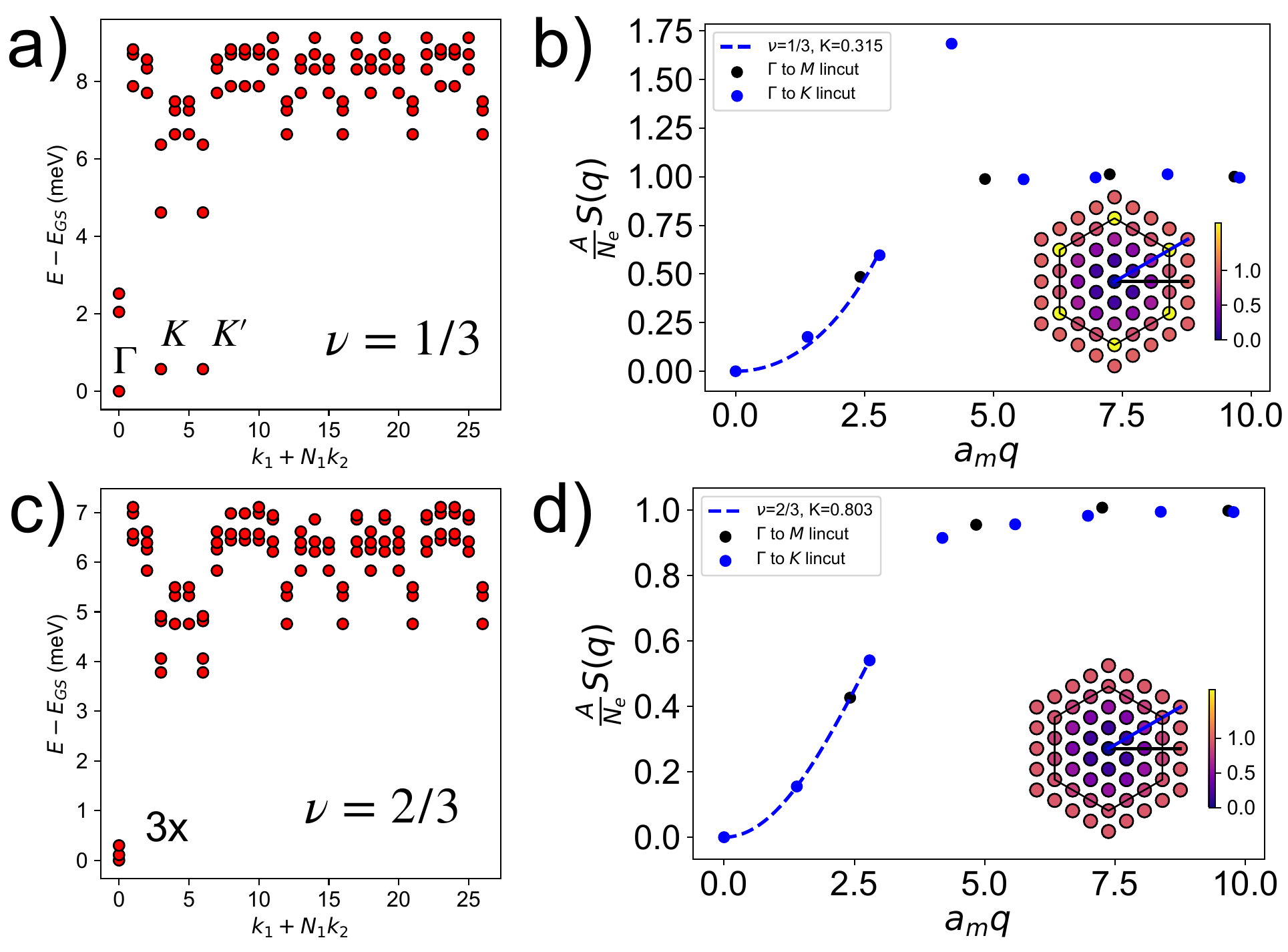} \quad \\\vskip 0.5cm
    \caption{Many Body Spectra ((a),(c)) and Structure Factor ((b),(d)) at $\theta=2.6$, $\epsilon=5$ for $\nu=1/3,2/3$ respectively.}
    \label{figS1: ED Data}
\end{figure*}

We reinforce the many-body neural Bloch wave function results and provide further support for the CDW and FQAH liquid phase at $\nu=1/3,2/3$ respectively. As seen in Fig. \ref{figS1: ED Data}, the many body spectra (MBS) shows 3 degenerate ground states at $\Gamma$, $K$, $K'$ in (a) and just at $\Gamma$ in (c). In Fig. \ref{figS1: ED Data} (a), $K$, $K'$ fold back to $\Gamma$ in the symmetry broken Brillouin zone as a result of the tripled unit cell of the 27 site cluster (ED cluster sizes and properties detailed in \cite{wilhelm2021interplay, reddy2023fractional}), indicative of CDW order.  Fig. \ref{figS1: ED Data} (b) provides further evidence for CDW phase by evincing the presence of a Bragg peak at the corresponding $K,K'$ points in the Brillouin zone. Contrastingly, Fig. \ref{figS1: ED Data} (c) reveals the 3 degenerate ground states at $\Gamma$ necessary for a fractional quantum anomalous Hall state. Fig. \ref{figS1: ED Data} (d) shows a consistent FQAH liquid signature with the absence of Bragg peaks and a quantum weight above the universal topological bound of 2/3 \cite{zaklama2024structure}; in contrast to the $\nu=1/3$ case, the topological lower bound is violated, with the quantum weight $K=0.315<1/3$. 

Fig. \ref{figS1: ED Data} (c), (d) are in good agreement with the structure factor results from the many-body neural Bloch-wavefunction shown in Fig. \ref{Fig:sq}, demonstrating signature consistent with a CDW at $\nu=1/3$ and an FQAH liquid at $\nu=2/3$. Further, it is interesting to note that even when Bragg peak scaling is taken into account, the Bragg peak for the CDW state, as calculated by the many-body neural Bloch-wavefunction, is larger than that calculated by ED. The heightened Bragg peak reflects the higher degree of electron crystallization in the CDW state, reflecting the additional electron correlation effects taken into account by the many-body neural Bloch wavefunction. 

\section{II. Neural Network Architecture Details}

A message passing graph network starts with a graph representation of the particle configurations as a graph $G^{0}=\{V^{0},E^{0}\}$ with node feature $V^0_i$ and edge feature $E^0_{ij}$. We utilize the message passing neural network based on the spirit in Ref.~\cite{pescia2023message} with a few features added to generate neural orbital transformation and neural backflow. 

The particle position and layer spin information are first initialized into a graph by defining a set of vertex vectors $V_i^{0}$ and edge vectors $E_{ij}^{0}$:
\begin{equation}
    V^{0}_i = (v^{0}_i(r_i, l_i),  h^0_i) \quad, \quad  E^{0}_{ij} = (e^{0}_{ij}(r_{ij}, l_i),  h^0_{ij}) 
\end{equation}

where each vertex and edge correspond to a particle and a pair of particles with indices $i$ and $j$ running from $1$ to $N$. $r_i$ is the $i$-th particle position, $r_{ij} = r_{i} - r_{j} $ is the displacement between particles $i$ and $j$, and $l_i=\pm 1$ for layer spin up and spin down respectively. $(\text{sin}(g r_i),\text{cos}(g r_i))$ with reciprocal momentum $g$ as initial feature for $v^{0}_{i}$, and $(\text{sin}(k (r_i - r_j))$, and $\text{cos}(k (r_i - r_j)))$ with mesh momentum $k$ is used as initial feature for $e^{0}_{ij}$. $h^0_i$ and $h^0_{ij}$ are randomly initialized parameters such that $V^0_i$ and $E^0_{ij}$ have feature dimensions both equal to 32.

The vertex and edge vectors are processed iteratively to generate $V^l_i$ and $E^l_{ij}$. The $l$-th iteration ($l \geq 2$) is:
\begin{align}
    Q_{ij} &= W_Q^l E_{ij}^{l-1}, \quad K_{ij} = W_K^l E_{ij}^{l-1},\\
    \label{Eq:vertex}
    m_{ij}^{l} &= F^{l}(\sum_n  Q_{in}  K_{nj}) \odot G^{l}( E_{ij}^{l-1}),\\
    h_{i}^{l} &= U^{l}([\sum_j m_{ij}^{l}, V_{i}^{l-1}]) + h_{i}^{l-1}, \\
    h_{ij}^{l} &= H^{l}([ m_{ij}^l,  E_{ij}^{l-1}]) + h_{ij}^{l-1}, \\
    V^{l}_i &= (v^{0}_i,  h^l_i) \quad, \quad E^{l}_{ij} = (e^{0}_{ij},  h^l_{ij})
\end{align}
where $\otimes$ is the tensor product, $\odot$ is the element-wise multiplication, $F^{l},G^{l},H^{l}, U^{l}$ are one-layer fully-connected neural networks with hidden dimension 32 and GELU activation function. $W^l_Q$ and $W^l_K$ are embedding matrices and we use $l=2$ iteration.

To construct the neural backflow transformation, we start with the Bloch basis function

\begin{equation}
\varphi_{k}(r_i,l_i) = \sum_{g} u^{l_i}_{k g}e^{i(k+g)r_i}
\end{equation}
where $k$ and $g$ are the mesh momentum and the reciprocal lattice momentum. $u_{kg}^{l_i}$ is the Bloch function coefficient, which can be obtained from mean field calculation. The neural network output is generated by the many-body backflow, 
\begin{equation}
\bm{r}_i = r_i + WV^{t}_i
\end{equation}
where $W$ is a complex-valued linear transformation with output dimension 2.

To construct the neural orbital transformation $J_{k,i}(\rl)$, we apply a two-layer MLP with a complex-valued output $\text{MLP}(\{k\},V_i^t)$ on the message passing neural network output. The MLP is a two-layer neural network with hidden dimesion 32 and GELU activation function in the hidden layers.

The momentum transformation matrix is given by $M_{\lambda k}$, where the $\lambda=1,...,N$ with $N$ the number of particles. $k$ is the mesh momentum index and the number of $k$ is chosen to be the same number of the mesh points in the first band.

\section{III. Neural Network Optimization Details}

\begin{figure*}[t]
    \centering
    \includegraphics[width=0.9\textwidth]{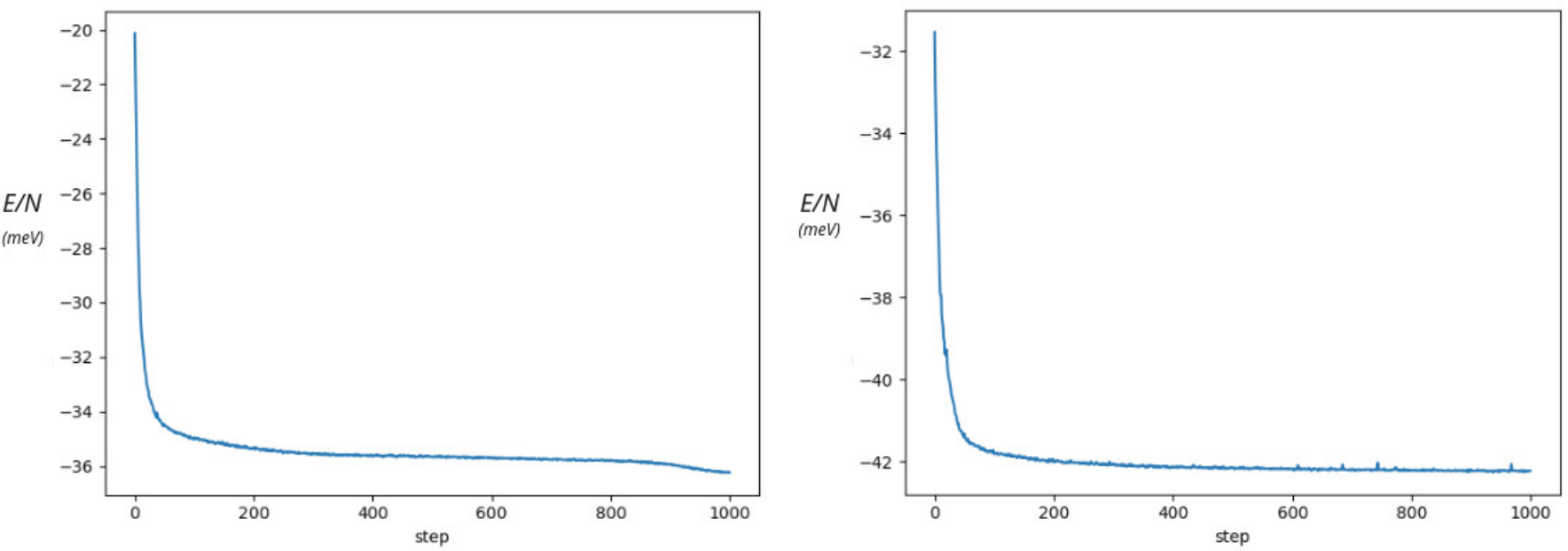} 
    \caption{NN energy optimization curve at $\nu=\frac{1}{3}$ and $\frac{2}{3}$ at $\theta=2.6$ on a 36 unit cells.}
    \label{Fig:opt}
\end{figure*}

We use natural gradient method~\cite{sorella1998green} to optimize the neural network. The learning rate is $\text{lr}=2 \times 10^{-3}$. The optimization takes 1000 steps for any clusters larger than $3 \times 3$. The number of MCMC samples during optimization is 4128. The initialization of the neural network uses LecunNormal~\cite{klambauer2017self}. The optimization is implemented in JAX~\cite{jax2018github}. We show the energy optimization curve of NN in Fig.~\ref{Fig:opt} as an example.

\end{document}